\documentclass[aps,prd,twocolumn,showpacs,preprintnumbers,nofootinbib,amsmath,amssymb]{revtex4}

\usepackage{graphicx}
\usepackage{bm}
\usepackage{amsthm,url}

\newcommand{\C}{\mathbb{C}}
\newcommand{\R}{\mathbb{R}}

\renewcommand{\vec}[1]{\mathbf{#1}}
\newcommand{\stack}[1]{\mathrm{vec}{(#1)}}

\newcommand{\cond}[1]{\mathrm{cond}\:{#1}}
\newcommand{\lag}{\langle}
\newcommand{\rag}{\rangle}

\newcommand{\diag}{\text{diag}}
\def\be{\begin{equation}}
\def\ee{\end{equation}}
\def\bi{\begin{itemize}}
\def\ei{\end{itemize}}
\def\bea{\begin{eqnarray}}
\def\eea{\end{eqnarray}}
\def\ba{\begin{array}}
\def\ea{\end{array}}
\def\no{\nonumber}
\def\bn{\begin{enumerate}}
\def\en{\end{enumerate}}
\def\p{\phi}
\def\t{\theta}
\def\a{\alpha}
\def\b{\beta}
\def\g{\gamma}
\def\e{\epsilon}

\begin{document}


\preprint{AEI-2007-119}

\title{Best network chirplet-chain: Near-optimal coherent detection
  of unmodeled gravitation wave chirps with a network of detectors} 
\author{Archana Pai}
\email{Archana.Pai@aei.mpg.de}
\affiliation{Max-Planck Institut f\"ur Gravitationsphysik, Am M\"uhlenberg 1, 14476 Potsdam, Germany}
\author{\'Eric Chassande-Mottin}
\email{Eric.Chassande-Mottin@apc.univ-paris7.fr}
\affiliation{CNRS, AstroParticule et Cosmologie, 10, rue Alice Domon et L\'eonie Duquet, 
75205 Paris Cedex 13, France}
\affiliation{Observatoire de la C\^ote d'Azur, Bd de l'Observatoire, BP 4229, 06304 Nice, France}
\author{Olivier Rabaste}
\email{Olivier.Rabaste@apc.univ-paris7.fr}
\affiliation{CNRS, AstroParticule et Cosmologie, 10, rue Alice Domon et L\'eonie Duquet, 
75205 Paris Cedex 13, France}
\date{\today}

\begin{abstract}
  The searches of impulsive gravitational waves (GW) in the data of
  the ground-based interferometers focus essentially on two types of
  waveforms: short unmodeled bursts from supernova core collapses and
  frequency modulated signals (or chirps) from inspiralling compact
  binaries. There is room for other types of searches based on
  different models. Our objective is to fill this gap. More
  specifically, we are interested in GW chirps ``in general'', i.e.,
  with an arbitrary phase/frequency \textit{vs.} time evolution.
  These unmodeled GW chirps may be considered as the generic
  signature of orbiting or spinning sources. We expect the
  quasi-periodic nature of the waveform to be preserved independently
  of the physics which governs the source motion. Several methods have
  been introduced to address the detection of unmodeled chirps using
  the data of a single detector. Those include the best chirplet chain
  (BCC) algorithm introduced by the authors. In the next years,
  several detectors will be in operation. Improvements can be expected
  from the joint observation of a GW by multiple detectors and the
  coherent analysis of their data namely, a larger sight horizon and
  the more accurate estimation of the source location and the wave
  polarization angles.  Here, we present an extension of the BCC
  search to the multiple detector case. This work is based on the
  coherent analysis scheme proposed in the detection of inspiralling
  binary chirps. We revisit the derivation of the optimal statistic
  with a new formalism which allows the adaptation to the detection of
  unmodeled chirps.  The method amounts to searching for salient
  paths in the combined time-frequency representation of two {\it
    synthetic streams}. The latter are time-series which combine the
  data from each detector linearly in such a way that all the GW
  signatures received are added constructively. We give a proof of
  principle for the full sky blind search in a simplified situation
  which shows that the joint estimation of the source sky location and
  chirp frequency is possible.
\end{abstract}
                                

\maketitle

\section{Summary}

A large effort is underway to analyze the scientific data acquired
jointly by the long-baseline interferometric gravitational wave (GW)
detectors GEO 600, LIGO, TAMA and Virgo \cite{detectors}.  In this
paper, we contribute to the methodologies employed for this analysis,
and in particular for the detection of impulsive GW signals.

The current GW data analysis effort is targeted on two types of
impulsive GWs.  A first target is poorly known short bursts of GWs
with a duration in the hundredth of a millisecond range. The
astrophysically known sources of such GW bursts are supernovae core
collapses (or other similar cataclysmic phenomenon). The second target
is frequency modulated signals or \textit{chirps} radiated by
inspiralling binaries of compact objects (either neutron stars (NS) or
black holes (BH)). These chirp waveforms are well modeled and expected
to last for few seconds to few minutes in the detector bandwidth. Our
objective is to enlarge the signal range of impulsive GWs under
consideration and to fill this gap between these two types.  More
specifically, we are interested in the detection of unmodeled GW
chirps which last from few tens of milliseconds to few seconds in the
detector bandwidth. We shall detail in the next section the
astrophysical motivation for considering this kind of GWs.

Simultaneous observation and analysis of the jointly observed data
from different GW detectors has definite obvious benefits.  First and
foremost, a GW detection can get confirmed or vetoed out with such a
joint observation. Further, the detector response depends on the
position and orientation of the source and polarization of the wave.
For this reason, the joint observation by multiple detectors gives
access to physical parameters such as source location and
polarization. The use of multiple detectors also allows to enlarge the
observational horizon and sky coverage.

Till date, several methods have been proposed and implemented to
detect unmodeled GW chirps using a single detector which include the
signal track search (STS)
\cite{Anderson:1999pr}, the chirplet track search
\cite{candes07:_detec} and the best chirplet chain (BCC) search
proposed by the authors \cite{chassande-mottin06:_chirplet_chains}.
However, none of the above addresses the multiple detector case.
This requires the designing of specific algorithms which are
able to combine the information received by the different
detectors.

In practice, there are two approaches adopted to carry out network
analysis from many detectors, \textit{coincidence} and
\textit{coherent} approach. In coincidence approach, the data from
each detector is processed independently and only coincident trigger
events (in the arrival time and the parameter values) are retained and
compared.  On the other hand, in the coherent approach, the network as
a whole is treated as a single ``sensor'': the data from various
detectors is analyzed jointly and combined into a single {\it network
  statistic} which is tested for detection. In the literature, it has
been shown that the coherent approach performs better than the
coincidence approach for GW short bursts \cite{Arnaud:2003zq} as well
as GW chirps from coalescing binaries \cite{Mukhopadhyay:2006qj}. Indeed, the
signal phase information is preserved with the coherent approach,
whereas it is not with the coincidence approach which therefore leads
to some information loss. Thus, we wish to follow the coherent
approach for our analysis.

Another reason for this choice is that the coincidence method is not
adequate for unmodeled chirps.  A large number of parameters (in the
same order as the number of signal samples) is needed to characterize
their frequency evolution. A coincident detection occurs when the
parameter estimates obtained from the analysis of the individual
detector data match. Because of the noise fluctuation, the
occurrence of such a coincidence is very unlikely when the number of
parameters is large, unless the incoming GW has very large amplitude.
In this article, we therefore adopt the coherent method and propose
the \textit{coherent extension of the BCC algorithm}.


Coherent schemes have been already developed for the detection of inspiralling
binary chirps \cite{Pai:2000zt,Finn:2000hj}. Here, we revisit the work
presented in \cite{Pai:2000zt} with a new formalism. Comments in
footnotes link the results presented here with the ones of
\cite{Pai:2000zt}. We show that the new formalism presented here
helps to
understand the geometry of the problem and it is simple to establish
connections with earlier works.

The outline of the paper goes as follows. In
Sec.~\ref{Sec:chirpModel}, we present and motivate our model of an
arbitrary GW chirp. In Sec.~\ref{sec:response}, we describe the response of the
detector network to an incoming GW chirp. Further, we show that the
linear component of the signal model (parameters acting as scaling
factors and phase shifts, so called extrinsic parameters) can be factorized.
This factorization evidences that the signal space can be represented as the direct
product of two two-dimensional spaces i.e., the GW polarization plane and the chirp
plane. This representation forms the backbone of the coherent
detection scheme that follows in the subsequent section.

In Sec.~\ref{sec:interpretation}, based on a detailed geometrical
argument, we show that the above signal representation manifests the
possible degeneracy of the response. This degeneracy has been already
noticed and studied at length in the context of burst detection
\cite{klimenko05:_const,mohanty06:_variab,rakhmanov06:_rank_tikhon}.
We investigate this question in the specific context of chirps and
obtain similar results as were presented earlier in the literature.

In Sec.~\ref{sec:LR}, we obtain the expression of the network
statistic. Following the principles of the generalized likelihood
ratio test (GLRT), the statistic is obtained by maximizing the network
likelihood ratio over the set of unknown parameters. We perform this
maximization in two steps. We first treat the linear part of the
parameterization and show that such a maximization is nothing but a
{\it least square} problem over the extrinsic parameters. The solution
is obtained by projecting the data onto the signal space. We further
study the effect of the response degeneracy on the resulting parameter
estimates.

The projection onto the signal space is a combined projection onto the
GW polarization and chirp planes. The projection onto the first plane
generates two {\it synthetic streams} which can be viewed as the
output of ``virtual detectors''. The network statistic maximized over
the extrinsic parameters can be conveniently expressed in terms of the
processing of those streams. In practice, the synthetic streams
linearly combine the data from each detector in such a way that the GW
signature received by each detector is added constructively.  With
this rephrasing, the source location angles can be searched over
efficiently.

Along with the projection onto the GW polarization plane, we also
examine the projection onto its complement. While synthetic streams
concentrate the GW contents, the so called null streams produced this
way combines the data such that the GW signal is canceled out. The
null streams are useful to veto false triggers due to instrumental
artifacts (which don't have this cancellation property).  The null
streams we obtained here are identical to the ones presented earlier
in GW burst literature \cite{Guersel:1990bn,Wen:2005ui,chatterji06:_coher}.

In Sec.~\ref{sec:intrinsic}, we perform the second and final step of
the maximization of the network statistic over the chirp phase
function. This step is the difficult part of the problem. For the one
detector case, we have proposed an efficient method, the BCC algorithm
which addresses this question. We show that this scheme can be adapted
to the multiple detector case in a straightforward manner, hence refer to
this as \textit{Best Network CC} (BNCC).

Finally, Sec.~\ref{sec:discuss} presents a proof of principle of the
proposed method with a full-sky blind search in a simplified
situation.

\section{Generic GW chirps}
\label{Sec:chirpModel}

\subsection{Motivation}
Known observable GW sources e.g., stellar binary systems, accreting
stellar systems or rotating stars, commonly involve either orbiting or
spinning objects. It is not unreasonable to assume that the similar
holds true even for the unknown sources.

The GW emission is essentially powered by the source dynamics which
thus determines the shape of the emitted waveform. Under linearized
gravity and slow motion (i.e., the characteristic velocity is smaller than
the speed of light) approximation, the quadrup\^ole formula
\cite{flanagan05} predicts that the amplitude of emitted GW is
proportional to the second derivative of the quadrup\^ole moment of the
physical system. When the dominant part of the bulk motion follows an
orbital/rotational motion, the quadrup\^ole moment varies
quasi-periodically, and so is the GW.

The more the information we have about the GW signal, the easier/better is the
detection of its signature in the observations. Ideally, this requires
precise knowledge of the waveform, and consequently requires precise
knowledge of the dynamics. This is not always possible. In general,
predicting the dynamics of GW sources in the nearly relativistic
regime requires a large amount of effort. This task may get further
complicated if the underlying source model involves
magnetic couplings, mass accretion, density-pressure-entropy
gradients, anisotropic angular momentum distribution.

Here, we are interested in GW sources where the motion is orbital/rotational 
but the astrophysical dynamics is (totally or partially) unknown.
While our primary target is the unforeseen sources (this is why we
remain intentionally vague on the exact nature of the sources),
several identified candidates enter this category because their
dynamics is still not fully characterized.  These include (see
\cite{chassande-mottin06:_chirplet_chains} for more details and
references) binary mergers, quasi normal modes from young hot rotating
NS, spinning BH accreting from an orbiting disk.
As motivated before, following the argument of quadrup\^ole approximation,
the GW signature for such sources is not completely undetermined: it is
expected to be quasi-periodic, possibly frequency modulated GW ;
in brief, it is {\it a GW chirp}.  This is the basic motivation for
introducing a generic GW chirp model, as described in the next section.

\subsection{Generic GW chirp model}

In this section, we describe the salient features of the generic GW
chirp model used in this paper. We motivate the nature of GW polarization,
the regularity of its phase and frequency evolution.

\subsubsection{Relation between the polarizations}

The GW tensor (in transverse traceless (TT) gauge),
associated to the GW emitted from slow-motion, weak
gravity sources are mostly due
to variations of the mass moments (in contrast to current moments) and
can be expanded in terms of mass multipole moments as \cite{Thorne:1980ru}
\begin{equation}
h^{TT}(t) \propto \sum_{l=2}^\infty \sum_{m=-l}^l (\nabla \nabla Y^{lm})^{\rm STF} \frac{d^l}{{dt}^l} I^{lm}(t-r/c). \label{eq:hsph}
\end{equation}

Here, STF means ``symmetric transverse-tracefree'', $Y^{lm}$ are the
spherical harmonics, and $I^{lm}$ are the mass multipole moments.

We consider here isolated astrophysical systems with anisotropic mass
distributions (e.g., binaries, accreting systems, bar/fragmentation
instabilities) orbiting/rotating about a well-defined
axis\footnote{This condition can be relaxed to precessing systems
  provided that the precession is over time scales much longer than
  the observational time, typically of order of seconds.}.  It can be
shown that these systems emit GW predominantly in the $l=|m|=2$ mode.
Contributions from any other mass moments are negligibly small
\footnote{Recently, numerical relativity simulations
  \cite{Berti:2007fi} demonstrated that this is a fairly robust
  statement in the specific context of inspiralling BH binaries. The
  simulations show that BH binaries emit GW dominantly with $l=|m|=2$.
  However, as the mass-ratio decreases, higher multipoles get excited.
  A similar claim was also made in the context of quasi-normal modes
  produced in the ring-down after the merger of two BH, on the basis of
  a theoretical argument, see \cite{Flanagan:1997sx}, page 4538.}.
The pure-spin tensor harmonic $(\nabla \nabla Y^{lm})^{\rm STF}$ term
provides the GW polarization.  For $l=|m|=2$ mode, we have
\begin{equation}
(\nabla \nabla Y^{22})^{\rm STF} \propto (1 + \cos^2 \e) e_+ + 2i \cos \e \; e_\times.
\end{equation}

The tensors $e_+$ and $e_\times$ form a pair of independent and
linear-polarization GW tensors ($e_\times$ is rotated by and
angle $\pi/4$ with respect to $e_+$).
The orbital inclination angle $\e$ is the angle between the line of
sight to the source (in Earth's frame) and the angular momentum vector (or the
rotation axis) of the physical system, see Fig. \ref{fig:angle_def} (a). 
This shows that the emitted GW in the considered case
carries both GW polarizations.

The GW tensor is fully described as $h^{TT}(t) = h_+(t) e_+ +
h_\times(t) e_\times$. The phase shift between the two polarizations
$h_+$ and $h_\times$ arises from the $I^{lm}$ term which is
proportional to the moment of inertia tensor for $l=|m|=2$.  The
quadratic nature of the moment of inertia tensor introduces a phase
shift of $\pi/2$ between the two polarizations $h_+$ and $h_\times$.
This leads to the chirp model below
\begin{subequations}
\label{eq:h}
\bea
h_+(t) &=& A \frac{1+\cos^2\e}{2} \: \cos(\varphi(t - t_0) + \phi_0) \label{eq:hplus} \, ,\\
h_\times(t) &=&  A \cos\e \: \sin(\varphi(t - t_0) + \phi_0), \label{eq:hcross}
\eea
\end{subequations}
with $t_0 \leq t < t_0+T$ and $h_{+,\times}(t)=0$ outside this
interval. The phase $\phi_0$ is the signal phase at $t=t_0$.

Here, we assume the GW amplitude $A$ to be constant. This is
clearly an over-simplified case since we indeed expect an amplitude
modulation for real GW sources. However we wish here to keep the model
simple in order to focus the discussion on the aspects related
to the coherent analysis of data from multiple detectors. We
postpone the study of amplitude modulated GWs to future work.

The chirp model described in Eq.~(\ref{eq:h}) clearly depends upon
several unknown parameters (which need to be estimated from the data)
which include the amplitude $A$, the initial phase $\phi_0$, the
arrival time $t_0$ of the chirp and the inclination angle $\epsilon$.
As no precise assumption on the exact nature and dynamics of the GW
source is made, we consider the phase evolution function
$\varphi(\cdot)$ to be an unknown parameter of the model (\ref{eq:h})
as well.  Clearly, it is a more complicated parameter than the others
which are simply scalars. Just like any scalar parameter can be
constrained to a range of values (e.g., $A>0$), the phase function
$\varphi(\cdot)$ has to satisfy conditions to be physically realistic
which we describe in the next section.

\subsubsection{Smoothness of the phase evolution}
\label{sec:smooth_chirps}
As explained above, the chirp phase is directly related to the orbital
phase of the source. The regularity of the orbital phase can be
constrained by the physical arguments: the orbital phase and its
derivatives are continuous. The same applies to the chirp phase and
derivatives.

The detectors operate in a frequency window limited in the range from
few tenths of Hz to a kHz and they are essentially blind outside. This
restricts our interest to sources emitting in this frequency range,
which results in lower and upper limits on the chirp frequency
$\nu(t)\equiv (2\pi)^{-1} d\varphi/dt$ and thus on the variations of
the phase.

In addition, the variation of the frequency (the chirping rate) can be
connected to the rate at which the source loses its energy. For
isolated systems, this is clearly bounded. This argument motivates the
following bounds on the higher-order derivatives of the phase:
\begin{align}
\label{smooth_def}
\left|\frac{d \nu}{dt}\right|&\leq F',&\left|\frac{d^2 \nu}{dt^2}\right|&\leq F''.
\end{align}

In this sense, Eq.(\ref{smooth_def}) determines and strengthens the smoothness of the
phase/frequency evolution. This is the reason why we coined the term
``smooth GW chirp'' in \cite{chassande-mottin06:_chirplet_chains}. The
choice of the allowed upper bounds $F'$ and $F''$ may be
based on general astrophysical arguments on the GW source of interest.

\section{Response of a network of detectors to an incoming GW}
\label{sec:response}

After describing the chirp model in the previous section,
in this section we derive the response of a network of
interferometric ground based detectors with
arbitrary locations and orientations to an incoming
GW chirp. The first step towards this is to identify the
coordinate frames.

\subsection{Coordinate frames}

We follow the conventions of \cite{Pai:2000zt} and introduce three
coordinate frames, namely, the wave
frame, the Earth frame and the detector frame as given below,
see Fig. \ref{fig:angle_def}.  
\begin{itemize}
\item the {\it wave} frame $\vec{x}_w\equiv(x_w,y_w,z_w)$ is the
  frame associated to the incoming GW with positive $z_w$-direction
  along the incoming direction and $x_w-y_w$ plane corresponds to the
  plane of the polarization of the wave.
\item the {\it Earth} frame $\vec{x}_E\equiv(x_E ,y_E ,z_E )$ is the
  frame attached to the center of the Earth. The $x_E$ axis is
  radially pointing outwards from the Earth's center and the
  equatorial point that lies on the meridian passing through
  Greenwich, England.  The $z_E$ axis points radially outwards from
  the center of Earth to the North pole.  The $y_E$ axis is chosen to
  form a right-handed coordinate system with the $x_E$ and $z_E$ axes.
\item the {\it detector} frame $\vec{x}_d\equiv(x_d ,y_d ,z_d)$ is the
  frame attached to the individual detector. The $(x_d-y_d)$ plane
  contains the detector arms and is assumed to be tangent to the
  surface of the Earth. The $x_d$ axis bisects the angle between the
  detector's arms. The $z_d$ axis points towards the local zenith. The
  direction of the $y_d$ axis is chosen so that we get a right-handed
  coordinate system.
\end{itemize}
 
\begin{figure*}[!htb]
\includegraphics[width=1.5\columnwidth]{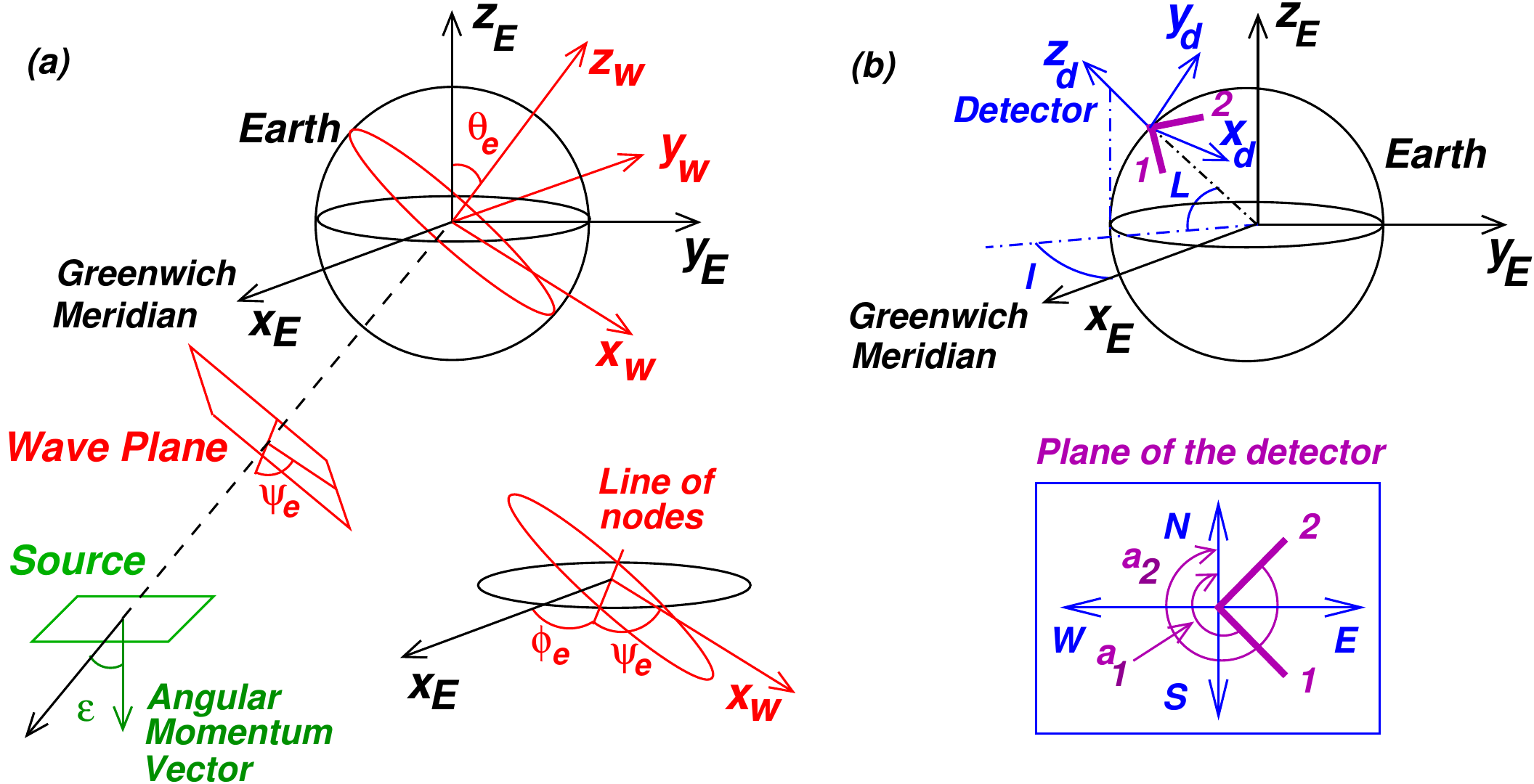}
\caption{\label{fig:angle_def} \textbf{Color online --- Coordinate
    transformations}.  (a) ${\cal O}(\p_e, \t_e, \psi_e)$: Earth frame
  $\vec{x}_E: (x_E,y_E,z_E) \rightarrow$ wave frame
  $\vec{x}_w:(x_w,y_w,z_w)$, (b) ${\cal O}(\a, \b, \g )$: Earth frame
  $\vec{x}_E:(x_E,y_E,z_E) \rightarrow $ detector frame:
  $\vec{x}_d:(x_d,y_d,z_d)$. The latitude $l$ and longitude $L$ of the
  detector are related to the Euler angles by Eqs. (\ref{eq:det_eul1})
  and (\ref{eq:det_eul2}).}
\end{figure*}

\begin{table*}
  \caption{\label{tab:det_euler} \textbf{Location and orientation informations of the Earth-based interferometric GW detectors.} The location of the corner station (vertex) of each detector is given in terms of the latitude and longitude. The orientation of an arm is given by the angle through which one must rotate it clockwise (while viewing from top) to point the local North. The corresponding detector Euler angles $(\a,\b,\g)$ are listed.}
\begin{center}
\begin{tabular}{lccccccc} 
Detector & Vertex & Vertex & Arm 1 & Arm 2 & $\a$ & $\b$ & $\g$ \\ 
& latitude (N) & longitude (E) & {$a_1$} &{$a_2$}   & & & \\ \hline 
TAMA-300 (T) & $35^\circ40'35.6''$ & $139^\circ 32'9.8''$  & $90.0^\circ$ & $180.0^\circ$ &$229.54^\circ$ & $54.32^\circ$ & $225^\circ$ \\
GEO-600 (G) & $52^\circ 14' 42.528''$  & $9^\circ 48'25.894''$   & $25^\circ 56' 35''$& $291^\circ 36' 42''$ &$99.81^\circ$ & $37.75^\circ$ & $68.778^\circ$ \\
VIRGO (V) & $43^\circ 37' 53.0921''$  & $10^\circ30'16.1878''$   & $341^\circ 25' 57.2''$ & $70^\circ 34'2.8''$ & $100.5^\circ$  & $46.37^\circ$ & $116.0^\circ$\\
LIGO Hanford (LH) & $46^\circ 27'18.528''$ & $-119^\circ 25'27.5657''$  & $35.9994^\circ$ & $125.9994^\circ$ & $-29.41^\circ$ & $43.55^\circ$ & $171.0^\circ$ \\
LIGO Livingstone (LL) & $30^\circ 33' 46.4196'' $ & $-90^\circ 46' 27.2654''$ & $107.7165^\circ $ & $197.7165^\circ$ & $-0.77^\circ$  &$59.44^\circ$& $242.72^\circ$ \\
\end{tabular}
\end{center}  
\end{table*}

A rotation transformation between the coordinate systems about the origin
is specified by the rotation operator ${\cal O}$ which is characterized
by these three Euler angles. We define these
angles using the ``$x$-convention'' (also known as $z-x-z$ convention)
\cite{goldstein80:_class}.

Let $(\p_e,\t_e,\psi_e)$ and $(\a,\b,\g)$ be Euler angles of the
rotation operator relating pairs of the above coordinate systems as follows
\bea
\vec{x}_w &=& {\cal O}(\p_e, \t_e, \psi_e) \vec{x}_E, \label{WtoE}\\
\vec{x}_d &=& {\cal O}(\a, \b, \g ) \vec{x}_E. \label{DtoE}
\eea

All the angles in the Eqs. (\ref{WtoE}) and (\ref{DtoE}) are related to
physical/geometrical quantities described in Fig. \ref{fig:angle_def}.
More specifically, we have
\begin{align}
\phi_e&= \phi-\pi/2,& \theta_e&= \pi-\theta,& \psi_e&=\psi,
\end{align}
where $\phi$ and $\theta$ are the spherical polar coordinates of the
source in the Earth's frame and the angle $\psi$ is the so-called
polarization-ellipse angle which gives the orientation of the source
plane. Throughout the paper, we shall use $\theta$ and $\phi$ to
indicate the source location.

The detector Euler angles $(\a,\b,\g)$ are directly related to the
location and orientation of the detector as follows:
\begin{align}
\label{eq:det_eul1}
\a &= L+\pi/2,\\
\label{eq:det_eul2}
\b &= \pi/2-l,\\
\g &= \frac{a_1+ a_2}{2}+ \frac{3 \pi}{2} & \text{if $|a_1-a_2| >   \pi$}, \\
   &= \frac{a_1+ a_2}{2}+ \frac{  \pi}{2} & \text{if $|a_1-a_2| \le \pi$},
\end{align}
where $l$ and $L$ are the latitude and longitude of the corner
station. The angles $a_1$ and $a_2$ describe the orientation of the
first and second arm respectively. It is the angle through which one
must rotate the arm clockwise (while viewing from top) to point the
local North.
In Table \ref{tab:det_euler}, we tabulate the currently running
interferometric detectors along with their Euler angles.

Combining Eqs. (\ref{WtoE}) and (\ref{DtoE}), we obtain
the coordinate transformation from the wave frame to the detector frame
as follows
\be
\vec{x}_s = {\cal O}(\p'_e, \t'_e, \psi'_e) \vec{x}_d \,,  
\ee
where ${\cal O}(\p'_e, \t'_e, \psi'_e) \equiv {\cal O}(\p_e, \t_e, \psi_e) {\cal O}^{-1}(\a,\b,\g)$.

\subsection{Network response}

An interferometric response to the incident GW is obtained by 
contracting a GW tensor with the detector tensor 
[see Appendix \ref{app:detresp}], which can be re-expressed
as a linear combination of the two polarizations $h_+$ and $h_\times$
{\it i.e.}
\bea
s&=& f_+ h_+ + f_\times h_\times \no \\
&\equiv& \Re[f^* h] \label{eq:gw_resp},
\eea
and the linear coefficients $f_+(\phi'_e,\theta'_e,\psi'_e)$ and
$f_\times(\phi'_e,\theta'_e,\psi'_e)$, commonly termed as the detector
antenna pattern functions, represent the detector's directional response
to the $+$ and $\times$ polarizations respectively. For the compact
expression provided by Eq.~(\ref{eq:gw_resp}), we have defined the
complex GW signal to be $h=h_+ + ih_\times$ and the complex antenna
pattern function to be $f = f_+ + i f_\times$.

The detector response $s$ and the incident GW signal $h$ are both
times series. In a network where the various detectors are located at
different locations on the Earth, for example the LIGO-Virgo network,
the GW will arrive at the detector sites at different time
instances. However, all the measurements at the various detectors need to be
carried out with a reference time. Here, just for our convenience, we choose
the observer attached to the Earth's center as a reference and
the time measured according to this observer is treated as a reference.
Any other reference would be equally acceptable.

Assuming that the detector response is labeled with this time reference
(a single reference time for all detectors in a detector network), we
have 
\be 
s(t)=\Re[f^* h(t-\tau(\phi,\theta))], 
\ee 
where $\tau(\phi,\theta)=(\vec{r}_d - \vec{r}_E) \cdot
\vec{w}(\phi,\theta)/c$ denotes the difference in the arrival times of
the GW (propagating with the unit wave vector $\vec{w}$) at the
detector and at the center of the Earth located at $\vec{r}_d$ and
$\vec{r}_E$, respectively. Note that this value can be positive or
negative depending on the source location.

\subsection{Vector formalism}
In the following, we distinguish scalars by using roman letters,
vectors are denoted by small bold letters, and matrices by bold
capitals. We denote the $k$-th element of vector $\vec{a}$ by $a[k]$
and correspondingly, the element of matrix $\vec{A}$ at row $k$ and
column $l$ by $A[k,l]$.  The matrices $\vec{A}^T$ and $\vec{A}^H\equiv
(\vec{A}^T)^*$ designate the real and hermitian transposes of
$\vec{A}$ respectively.

We consider now a GW detector network with $d$ interferometers. Each detector 
and its associated quantities are labeled with
an index $j =1, \dots ,d$ which we also use as a subscript if required.
We assume that the output response of each detector is sampled at the
Nyquist rate $\nu_s\equiv 1/t_s$ where $t_s$ is the sampling interval.
We then divide the data in blocks of
$N$ consecutive samples.  In this set-up, the detector as well as
the network response is then defined by forming vectors with these blocks
of data.

Let us consider a given GW chirp source at sky location
$(\phi,\theta)$. Let the response of the $j$-th detector be
$\vec{s}_j$ with entry $s_j[k]=s_j(t_k+\tau_j(\phi,\theta)) $, where
$t_k=t_0+k t_s$, $k=0,\ldots ,N-1$, and $t_0$ is the reference time
{\it i.e.} the time of arrival of GW at the center of the Earth.
Note that, with the above definition,
we compensate for the time delay $\tau_j(\phi,\theta)$ between the
detector $j$ and the Earth's center. Thus, in this set-up, the GW
signal starts and ends in the same rows in the data vectors
$\vec{s}_j$ of all the detectors.

For compactness, we stack the data from all the detectors in the
network into a single vector $\vec{s}$ of size $Nd \times 1$, such that
$\vec{s}^T=[\vec{s}_1^T \vec{s}_2^T \ldots \vec{s}_d^T]$ forms
the \textit{network response}. In this convention, the network
response can be expressed compactly
as the {\it Kronecker product} (see Appendix \ref{app:kron} for the
definition) of the network complex beam pattern vector $\vec{f} = \{
f_j, j=1\ldots d \} \in \C^{d \times 1}$ and the complex GW vector
$\vec{h} = \{ h(t_k), t_k=t_0+t_s k \mathrm{~with~} k=0\ldots N-1\}
\in \C^{N \times 1}$ viz.,
\be
\vec{s} = \Re[\vec{f}^* \otimes \vec{h}]. \label{eq:netS1}
\ee

The above expression is general enough to hold true for any type of
incoming GW signal. The Kronecker product in this expression is the
direct manifestation of the fact that the detector response is nothing
but the tensor product between the detector and the wave tensors.

\subsection{GW chirp as a linear model of the extrinsic parameters}
\label{sec:GW_linear}
In previous section, we have obtained the network response to any type of incoming GW
with two polarizations.
In what follows, we wish to investigate how this response manifests
in case of a specific type of GW, namely GW chirp described in Eq.~(\ref{eq:h}).
We also want to understand how various parameters explicitly appear
in the network response.


It is insightful to distinguish the signal parameters
based on their effect on the signal model.
The parameters are separated into two distinct types traditionally
referred to as the ``\textit{intrinsic}'' and ``\textit{extrinsic}''
parameters. The extrinsic parameters are those that introduce scaling
factors or phase shifts but do not affect the shape of the signal
model. Instead, intrinsic parameters significantly alter the shape of
the signal and hence the underlying geometry.

The network response $\vec{s}$ mingles these two types of parameters.
Our work is considerably simplified if we can ``factorize'' the
extrinsic parameters from the rest. For the chirp model described in
Eq.~(\ref{eq:h}), we count four extrinsic parameters, namely
$\{A,\phi_0,\e,\psi_e\}$ and perform this factorization in two steps.

\subsubsection{Extended antenna pattern includes the inclination angle}
\label{sec:extendedAP}
We absorb the inclination angle $\epsilon$ into the 
antenna pattern functions and rewrite the network signal as
\be
\vec{s} = \Re[\vec{\tilde{f}}^* \otimes \vec{\tilde{h}}], \label{eq:netS2}
\ee
where $\vec{\tilde{h}} \equiv a \vec{e}$ is the
GW vector. It 
only depends on the complex amplitude $a=A \exp i\phi_0$ and on
the phase vector $\vec{e}=\{\exp(i \varphi[k]), \text{ with }
\varphi[k]\equiv \varphi(k t_s), k=0\ldots N-1\}$.

The {\it extended} antenna pattern $\vec{\tilde{f}}$ incorporates the
inclination angle $\e$ as follows\footnote{We remind the reader that
  similar quantity was previously introduced in Eq.~(3.19) of
  \cite{Pai:2000zt}.}
\be
\label{eq:ext_ant}
\vec{\tilde{f}} = \frac{1 + \cos^2 \e}{2}~\vec{f}_+ + i \cos \e~\vec{f}_{\times}\,.
\ee

\subsubsection{Gel'fand functions factorize the polarization angles from source location angles}

The second step is to separate the dependency of $\vec{\tilde{f}}$ on
the polarization angles $\{\psi,\epsilon\}$ from
the source location angle and the detector orientation angles.
The earlier work \cite{Pai:2000zt} shows that the Gel'fand
functions (which are representation of the rotation group $SO(3)$)
provide an efficient tool to do the same.
For the sake of completeness, Appendix
\ref{app:detresp} reproduces some of the calculations of
\cite{Pai:2000zt}. The final result (see also Eqs.~(3.14-3.16) of
\cite{Pai:2000zt}) yields the following decomposition:
\be
\label{eq:extF}
\vec{\tilde{f}} = t_+ \vec{d} + t_- \vec{d}^*,
\ee
where the vector $\vec{d} \in \C^{d \times 1}$ carries the information
of the source location angles $(\phi,\theta)$ {\it via} $(\phi_e,\t_e)$
and the detector Euler angles $\{\a_j,\b_j,\g_j\}$. Its components are expressed as
\be
d[j]=- \sum_{n=-2}^{n=2} i T_{2 n}(\phi_e,\theta_e,0)[T_{2 n}(\a_j,\b_j\g_j) - T_{-2 n} (\a_j,\b_j,\g_j)]^*\, .
\ee
The tensor
$T_{mn}$ designates rank-2 Gel'fand functions.
The coefficients $t_+$ and $t_-$ depend only on
the polarization angles $\{\psi,\epsilon\}$, viz.
\be
\label{eq:tplusminus}
t_\pm = T_{2 \pm 2}(\psi,\e,0) = \frac{(1 \pm \cos \e)^2}{4} \exp(\mp 2 i \psi).
\ee

Finally, we combine Eqs. (\ref{eq:netS2}) and (\ref{eq:extF}) and
obtain an expression of the network response where the extrinsic
parameters are ``factorized'' as follows,
\be
\vec{s} = \left(\underbrace{\left[\ba{cc} {\vec d} & {\vec d^*} \ea \right]}_{\vec{D}} \otimes \underbrace{\frac{1}{2}\left[\ba{cc} {\vec{e}} & {\vec{e}^*} \ea \right]}_{\vec{E}} \right) \underbrace{\left(\ba{c} a~t^*_-  \\  a^*t_+ \\ a~t^*_+ \\ a^*t_-\ea\right)}_{\vec{p}} \equiv \vec{\Pi} {\vec{p}}.  \label{eq:netSig}
\ee

Eq.~(\ref{eq:netSig}) evidences the underlying linearity of the GW
model with respect to the extrinsic parameters. The 4-dimensional
complex vector $\vec{p}$ defines a one-to-one (non-linear) mapping
between its components and the four physical extrinsic parameters
$\{A,\phi_0,\e,\psi\}$ (we will detail this point later in Sec.
\ref{sec:est_extrinsic}).  Note that the first and fourth components
as well as the second and third components of $\vec{p}$ are complex
conjugates. This symmetry comes from the fact that the data is real.

The \textit{signal space} as defined by the network response
is the range of $\vec{\Pi}$ and results from
the Kronecker product of two linear spaces: the plane of $\C^d$
generated by the columns of $\vec{D}$ which we shall refer to as
\textit{GW polarization plane} \footnote{In \cite{Pai:2000zt}, this plane was
  referred to as ``helicity plane'' because it is formed by the
  network beam patterns for all possible polarizations.} and the plane
of $\C^N$ generated by the columns of $\vec{E}$ which we shall refer
to as \textit{chirp plane}. These two spaces embody two fundamental
characteristics of the signal: the former characterizes gravitational
waves while the latter characterizes chirping signals.  The Kronecker
product in the expression of $\vec{\Pi}$ shows explicitly that the
network response is the result of the \textit{projection} of incoming
GW onto the detector network.

The norm of the network signal gives the ``signal'' (and not physical)
energy delivered to the network, which is
\be \big
\|\vec{s} \big \|^2 = \frac{N A^2} {2} \big \| \vec{\tilde{f}} \big\|^2 \,.
\label{eq:rho}
\ee

Clearly, the dependence on the number of samples $N$ implies that the
longer the signal duration, the larger is the signal energy and is
proportional to the length of the signal duration. The factor
$\big \| \vec{\tilde{f}} \big \|$ is the modulus of the extended antenna
pattern vector. It can be interpreted as the gain or attenuation
depending on the direction of the source and on the polarization of
the wave.

\section{Interpretation of the network response}
\label{sec:interpretation}
In this section, we focus on understanding the underlying geometry of
the signal model described in Eq.~(\ref{eq:netSig}). A useful tool to
do so is the Singular Value Decomposition (SVD)
\cite{golub96:_matrix}.  It provides an insight on the geometry by
identifying the principal directions of linear transforms.

\subsection{Principal directions of the signal space: Singular Value Decomposition}

The SVD is a generalization of the eigen-decomposition for non-square
matrices. The SVD factorizes a matrix $\vec{A} \in \C^{m \times n}$
into a product $\vec{A} = \vec{U}_A \vec{\Sigma}_A \vec{V}^H_A$ of
three matrices $\vec{U}_A \in \C^{m \times r}$, $\vec{\Sigma}_A \in
\R^{r \times r}$ and $\vec{V}_A \in \C^{n \times r}$ where $r \le m,n$
is the rank of $\vec{A}$. The columns of $\vec{U}_A$ and $\vec{V}_A$
are orthonormal i.e., $\vec{U}_A^H \vec{U}_A=\vec{V}_A^H
\vec{V}_A=\vec{I}_r$. The diagonal of $\vec{\Sigma}_A$ are the singular
values (SV) of $\vec{A}$. We use here the so-called ``compact'' SVD
(we retain the non-zero SV only in the decomposition), such that the 
matrix $\vec{\Sigma}_A$ is a positive definite diagonal matrix.

The SVD is compatible with the Kronecker product
\cite{horn91:_topic_matrix_analy}: the SVD of a Kronecker product is
the Kronecker product of the SVDs. Applying this property to
$\vec{\Pi}$, we get 
\be
\label{eq:svdpi}
\vec{\Pi} = (\vec{U}_D \otimes \vec{U}_E) (\vec{\Sigma}_D \otimes \vec{\Sigma}_E) (\vec{V}_D \otimes \vec{V}_E)^H.
\ee

Therefore, the SVD of $\vec{\Pi}$ can be easily deduced from the one of
$\vec{D}$ and $\vec{E}$. We note that $\vec{D}$ and $\vec{E}$ have similar
structure (two complex conjugated columns), see Eq. (\ref{eq:netSig}).
In Appendix \ref{app:svd_rank2},
we analytically obtain the SVD of a matrix with such a structure.
Thus, applying this result, we can straightaway
write down the SVDs for $\vec{D}$ and $\vec{E}$ as shown in
the following sections.

\subsubsection{GW polarization plane: SVD of $\vec{D}$}
\label{sebsec:hel}
Let us first introduce some variables
\bea
{\cal D} &\equiv& \vec{d}^H \vec{d} = \sum_{j=1}^d |d[j]|^2 \,, \\
\Delta &\equiv& \vec{d}^T \vec{d} = \sum_{j=1}^d d[j]^2 \,, \\
\delta &\equiv& \arg \Delta.
\eea

In the nominal case, the matrix $\vec{D}$ has rank 2, viz.
\be
\vec{\Sigma}_D = \left[\ba{cc} \sigma_1 & 0\\
0 & \sigma_2 \ea\right],
\ee
with two non-zero SV $\sigma_1=\sqrt{{\cal D} + |\Delta|}$ and $\sigma_2=\sqrt{{\cal D} - |\Delta|}$ ($\sigma_1 \ge \sigma_2$)
associated to a pair of left-singular vectors $\vec{V}_D = [\vec{v}_1, \vec{v}_2]$ with
\begin{align}
\label{eq:vD}
\vec{v}_1&=\frac{1}{\sqrt{2}}\left[\ba{c}\exp(-i \delta)\\ 1\ea\right] \,, ~~~&
\vec{v}_2&=\frac{1}{\sqrt{2}}\left[\ba{c}\exp(-i \delta)\\ -1\ea\right] \,,
\end{align}
and of right-singular vectors $\vec{U}_D = [\vec{u}_1, \vec{u}_2]$ with
\bea
\label{eq:u1}
\vec{u}_1&=&\frac{\exp(-i \delta) \vec{d} + \vec{d}^*} {\sqrt{2({\cal D}+|\Delta|)}} \,,\\
\label{eq:u2}
\vec{u}_2&=&\frac{\exp(-i \delta) \vec{d} - \vec{d}^*} {\sqrt{2({\cal D}-|\Delta|)}}.
\eea

Note that the vector pair $\{\vec{u}_1, \vec{u}_2\}$ results from the
Gram-Schmidt ortho-normalization of $\{\vec{d}, \vec{d}^*\}$.

Barring the nominal case, for a typical network built with the
existing detectors and for certain sky locations of the source, it is
however possible for the smallest SV $\sigma_2$ to vanish. In such
situation, the rank of $\vec{D}$ reduces to 1. We then have
$\vec{\Sigma}_D = \sigma_1$, $\vec{V}_D = \vec{v}_1$ and $\vec{U}_D =
\vec{u}_1$. We give an interpretation of this degeneracy later in
Sec.~\ref{sec:rank_degen}.

\subsubsection{Chirp plane: SVD of $\vec{E}$}
The results of the previous section essentially apply to SVD calculation
of $\vec{E}$. However, there is an additional simplification due to
the nature of the columns of $\vec{E}$. Indeed, the cross-product
\be
\vec{e}^T \vec{e} = \sum_{k=0}^{N-1} \exp(2 i \varphi[k]) \,,
\ee
is an oscillating sum.  This sum can be shown
\cite{chassande-mottin06:_chirplet_chains} to be of small amplitude
under mild conditions compatible with the case of
interest. We can thus consider \footnote{This amounts to saying that
  the two GW polarizations (i.e., the real and imaginary parts of
  $\exp i\varphi[k]$) are orthogonal and of equal norm.  Note that this
  approximation is not required and can be relaxed. This would lead
  to use a version of the polarization pair ortho-normalized with a
  Gram-Schmidt procedure.} that $\vec{e}^T \vec{e} \approx 0$ and
$\vec{e}^H \vec{e} = N$.  Therefore, following Appendix \ref{app:svd_rank2} 2, 
the SVD of $\vec{E}$ is
given by $\vec{\Sigma}_E = \sqrt{N}\,\vec{I}_2/2$, $\vec{V}_E
=\vec{I}_2$ and $\vec{U}_E = 2\vec{E}/\sqrt{N}$.

\subsubsection{Signal space: SVD of $\vec{\Pi}$}
We obtain the SVD for $\vec{\Pi}$ using the compatibility of the SVD
with the Kronecker product stated in Eq.~(\ref{eq:svdpi}). In the
nominal case where $\vec{D}$ has rank 2, we have
\be
\label{eq:sigPi}
\vec{\Sigma}_\Pi = \frac{\sqrt{N}}{2} \left[\ba{cc} \sigma_1 \vec{I}_2 & \vec{0}_2 \\
\vec{0}_2 & \sigma_2 \vec{I}_2 \ea\right],
\ee
with four left-singular vectors
\be
\label{eq:Vpi}
\vec{V}_\Pi = \left[ \ba{ccc} \vec{v}_1 \otimes \vec{I}_2 & &\vec{v}_2 \otimes \vec{I}_2 \ea \right],
\ee
and four right-singular vectors
\be
\label{eq:Upi}
\vec{U}_\Pi = \frac{2}{\sqrt{N}} \left[\ba{ccccccc} \vec{u}_1 \otimes \vec{e} & &  \vec{u}_1 \otimes \vec{e}^* & & \vec{u}_2 \otimes \vec{e} & & \vec{u}_2 \otimes \vec{e}^* \ea \right].
\ee

\subsection{The signal model can be ill-posed}
\label{sec:rank_degen}
In the previous section, we obtained the SVD of $\vec{\Pi}$ in the
nominal case where the matrix $\vec{D}$ has 2 non-zero SVs. As we have
already mentioned, for a typical detector network, there might exist certain
sky locations where the second SV $\sigma_2$ of $\vec{D}$ vanishes
which implies that the rank of $\vec{D}$ degenerates to 1.  In such
cases, this degeneracy propagates to $\vec{\Pi}$ and subsequently its
rank reduces from 4 to 2.

In order to realize the consequences of this degeneracy, we first
consider a network of \textit{ideal} GW detectors (with no
instrumental noise).  Let a GW chirp pass through such a network from
a source in a sky location where $\sigma_2=0$. The detector output is
\textit{exactly} equal to $\vec{s}$. An estimate of the source
parameters would then be obtained from the network data by inverting
Eq.~(\ref{eq:netSig}).  However, in this case, this is impossible
since it requires the inversion of an under-determined linear system
(there are 4 unknowns and only 2 equations).

This problem is identical to the one identified and discussed at
length in a series of articles devoted to unmodeled GW bursts
\cite{klimenko05:_const,mohanty06:_variab,rakhmanov06:_rank_tikhon},
where this problem is formulated as follows: at those sky locations where
$\vec{D}$ is degenerated, the GW response is essentially made of only
\textit{one} linear combination of the \textit{two} GW polarizations.
It is thus impossible to separate the two individual polarizations
(unless additional prior information is provided). We want to stress
here that this problem is not restricted to unmodeled GW bursts but
also affects the case of chirping signals (and extends to the chirps
from inspiralling binaries of NS or BH \footnote{Contrary to the generic chirp model considered here,
  the phase and amplitude functions of inspiralling binary chirps
  follow a prescribed power-law time evolution. These differences
  affect only the geometry of the ``chirp plane'', but not that of the
  ``GW polarization plane'', hence the conclusion on the degeneracy remains
  the same.}). This is mainly because the degeneracy arises from the
geometry of the GW polarization plane which is same for any type of
source.

The degeneracy disappears at locations where $\sigma_2>0$ even if it
is infinitesimally small. However, when $\sigma_2$ is small, the
inversion of the linear equations in Eq.~(\ref{eq:netSig}) is very
sensitive to perturbations. With \textit{real world} GW detectors,
instrumental noise affects the detector response i.e., perturbs the
left-hand side of Eq.~(\ref{eq:netSig}).

A useful tool to investigate this is the {\it condition number}
\cite{rakhmanov06:_rank_tikhon}. It is a well-known measure of the
sensitivity of linear systems. The condition number of a matrix
$\vec{A}$ is defined as the ratio of its largest SV to the smallest.
For unitary matrices, $\cond(\vec{A})=1$. On the contrary, if $\vec{A}$ is
rank deficient, $\cond(\vec{A}) \rightarrow \infty$.  For the matrix
$\vec{\Pi}$, we have
\begin{equation}
\cond(\vec{\Pi}) = \frac{\sigma_1}{\sigma_2} = \sqrt{\frac{{\cal D}+|\Delta|}{{\cal
      D}-|\Delta|}}.
\end{equation}

\begin{figure}
\includegraphics[width=\columnwidth]{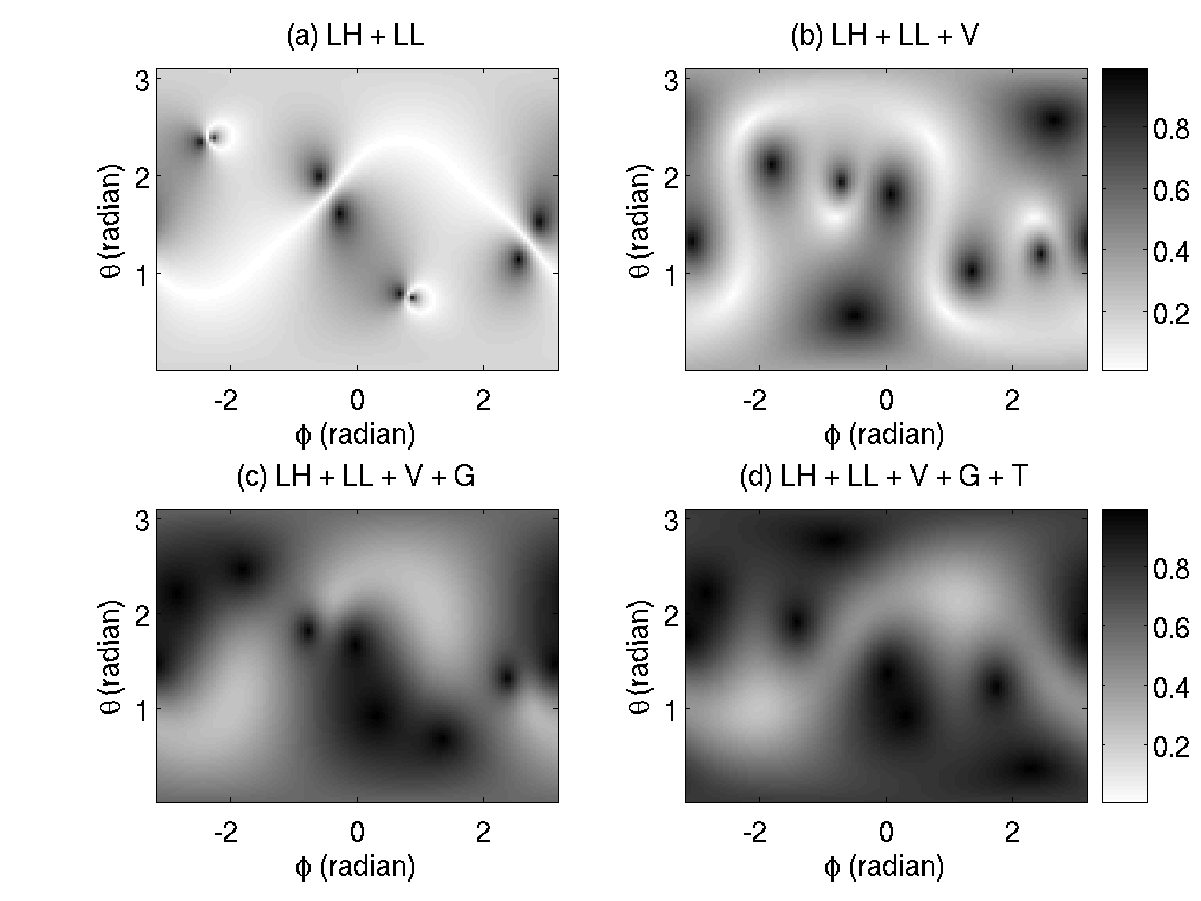}
\caption{\label{fig:cond_network} \textbf{Degeneracy
    of the network response}.  We show here the inverse of
  $\cond(\vec{\Pi})$ for various detector networks (the abbreviated
  detector names are listed in Table \ref{tab:det_euler}). The
  brighter regions of the sky correspond to the large conditioning
  number $\cond(\vec{\Pi})$.  The fraction of the sky where
  $1/\cond(\vec{\Pi}) < 0.1$ is (a) $25 \%$ (b) $4.5 \%$ (c) $\sim 0
  \%$ (d) $\sim 0 \%$. Since the LIGO detectors are almost aligned and
  they show the largest percentage of degeneracy.}
\end{figure}

In Fig. \ref{fig:cond_network}, we show full-sky plots of
$1/\cond(\vec{\Pi})$ for various configurations (these figures
essentially reproduce the ones of \cite{rakhmanov06:_rank_tikhon}).
We see that, even for networks of misaligned detectors, there are
significantly large patches where $\cond(\vec{\Pi})$ takes large
values. In those regions, the inversion of Eq.~(\ref{eq:netSig}) is
sensitive to the presence of noise and the estimate of the extrinsic
parameters thus have a large variance.

\paragraph{Connection to the antenna pattern function} ---
Interestingly, the SVs of $\vec{D}$ and $\vec{F}\equiv [\vec{f}
\: \vec{f}^*]$ coincide. This can be seen from the following
relationships we directly obtained from the definitions in Eqs.
(\ref{eq:ext_ant}), (\ref{eq:extF}) and (\ref{eq:tplusminus})
\begin{equation}
\begin{bmatrix}\tilde{\vec{f}}&\tilde{\vec{f}}^*\end{bmatrix}=
\vec{D} \begin{bmatrix} t_+&t_-^*\\t_-&t_+^*\end{bmatrix}=
\vec{F} \begin{bmatrix} |t_+|&|t_-|\\|t_-|&|t_+|\end{bmatrix}.
\end{equation}

The matrix $\vec{F}$ can be obtained from $\vec{D}$ by a unitary
transformation. Both matrices share the same singular spectrum.  We
can therefore write
\begin{equation}
\sigma^2_2 = \sum_{j=1}^d |f[j]|^2 - \left|\sum_{j=1}^d f[j]\right|^2.
\end{equation}

When $\sigma_2 \sim 0$, we thus have $|\vec{f}_+ \times \vec{f}_\times |^2
\sim 0$ where $\vec{f}_+\equiv\Re[\vec{f}]$ and
$\vec{f}_\times\equiv\Im[\vec{f}]$ are the network antenna pattern
vectors. This means that at such sky
locations, the antenna pattern vectors get aligned even if the
detectors in the network are misaligned. In other words, despite the
considered network is composed of misaligned detectors, it acts as a
network of aligned detectors at those sky locations. (Of course, for
perfectly co-aligned detectors, $\vec{f}_+ \propto \vec{f}_\times$ at
all sky locations.) Networks with many different detectors having
different orientations are
less likely to be degenerate.  This is confirmed in Fig.
\ref{fig:cond_network} where we see that the size of the degenerate
sky patches reduces when the number of detector with varied
orientations increases. For instance, with a network of LL-LH-G-V,
(assuming they have identical noise spectrum), it goes to zero.

\section{Network likelihood analysis: GW polarization plane and synthetic streams}
\label{sec:LR}
Generally speaking, a signal detection problem amounts to testing the
null hypothesis ($H_0$) (absence of signal in the data) {\it vs} the
alternate hypothesis ($H_1$) (presence of signal in the data).  Due to
the presence of noise, two types of errors occur: false dismissals (decide
$H_0$ when $H_1$ is present) and false alarms (decide $H_1$ when $H_0$
is present). There exist several objective criteria to determine the
detection procedure (or statistic) which optimizes the occurrence of
these errors. We choose the Neyman-Pearson (NP) approach which
minimizes the number of false dismissals for a fixed false alarm rate.
It is easily shown that for simple problems, the likelihood ratio (LR)
is NP optimal.  However, when the signal depends upon unknown
parameters, the NP optimal (uniformly over all allowed parameter
values) statistic is not easy to obtain. Indeed for most real-world
problems, it does not even exist. However, the generalized likelihood
ratio test (GLRT) \cite{Kay:1998} have shown to give sensible results
and hence is widely used.  In the GLRT approach, the parameters are
replaced by their maximum likelihood estimates. In other words, the
GLRT approach uses the maximum likelihood ratio as the statistic.
Here, we opt for such a solution.

As a first step, we consider the simplified situation where all
detectors have independent and identical instrumental noises and this
noise is white and Gaussian with unit variance. We will
address the colored noise case later in Sec. \ref{sec:color}.

In this case, the logarithm of the network likelihood ratio
(LLR) is given by
\be
\Lambda(\vec{x}) = -\big\| \vec{x} - \vec{s} \big\|^2 + \big\| \vec{x} \big\|^2, \label{eq:mlr}
\ee  
where $\big\|\cdot\big\|^2$ is the Euclidean norm (here in $\R^{Nd}$)
and we omitted an unimportant factor $1/2$. The network data vector
$\vec{x}$ is constructed on the similar lines as that of the
network response $\vec{s}$,
{\it i.e.} first, it stacks the data from all the detectors into
$\vec{x}^T=[\vec{x}_1^T, \vec{x}_2^T, \ldots \vec{x}_d^T]$ and then at
each detector, the data is time-shifted to account for the delay in
the arrival time $\vec{x}_j=\{x_j[k]=x_j(t_k+\tau_j), t_k=t_s k
\mathrm{~~and~~} k=0\ldots N-1\}$.


\subsection{Maximization over extrinsic parameters: scaling factors and phase shifts}

Following the GLRT approach, we maximize the network LLR $\Lambda$
with respect to the parameters of $\vec{s}$.  We replace $\vec{s}$ by
its model as given in Eq.~(\ref{eq:netSig}) and consider at
first the maximization with respect to the
extrinsic parameters $\vec{p}$.

\subsubsection{Least-square fit}

The maximization of the network LLR over $\vec{p}$ amounts to fitting
a linear signal model to the data in {\it least square} (LS) sense,
viz.  \be
\label{eq:mlike}
\text{minimize} \quad -\Lambda(\vec{x}) + \big\|\vec{x}\big\|^2 = \big\| \vec{x} - \vec{\Pi} \vec{p} \big\|^2 \quad \text{over $\vec{p}$}.
\ee

This LS problem is easily solved using the pseudo-inverse
$\vec{\Pi}^\#$ of $\vec{\Pi}$ \cite{golub96:_matrix}. 
The estimate of $\vec{p}$ is then given by 
\be
\label{eq:lsq}
\hat{\vec{p}} = \vec{\Pi}^\# \vec{x} \,.
\ee

The pseudo-inverse can be expressed using the SVD of $\vec{\Pi}$
as $\vec{\Pi}^\#=\vec{V}_\Pi \vec{\Sigma}_\Pi^{-1} \vec{U}_\Pi^H$
(note that $\vec{\Pi}^\#$ is always defined since we use the
\textit{compact} SVD restricted to non-zero SVs).

Substituting Eq.~(\ref{eq:lsq}) in Eq.~(\ref{eq:mlike}), we get the LS minimum to be 
\be
\label{eq:lsq2}
-\hat{\Lambda}(\vec{x}) +  \big\|\vec{x}\big\|^2 = \big\| \vec{x} - \vec{U}_\Pi \vec{U}_\Pi^H \vec{x} \big\|^2,
\ee
where we used $\vec{V}_\Pi^H \vec{V}_\Pi=\vec{I}_r$. Eq. (\ref{eq:lsq2}) can be further simplified into
\footnote{For inspiral case, this expression is equivalent to Eq.~(4.8) of \cite{Pai:2000zt}.}
\be
\label{eq:lam}
\hat{\Lambda}(\vec{x}) = \big\| \vec{U}_\Pi^H \vec{x} \big\|^2.
\ee
It is interesting to note that the operator $\vec{U}_\Pi
\vec{U}_\Pi^H$ is a (orthogonal) projection operator onto the signal space
(over the range of $\vec{\Pi}$) {\it i.e.} $\vec{U}_\Pi \vec{U}_\Pi^H
\vec{\Pi}=(\vec{\Pi} \vec{\Pi}^\#) \vec{\Pi} =\vec{\Pi}$.

\subsubsection{Signal-to-noise ratio}
\label{sec:snr}
The signal-to-noise ratio (SNR) measures the level of difficulty for
detecting a signal in the noise. In the present case, along with the
amplitude and duration of the incoming GW, the network SNR also depends
on the relative position, orientation of the source with respect to the
network. Therefore, the SNR should incorporate all these aspects. A
systematic way to define the SNR is to start from the statistic.

Let the SNR $\rho$ of an injected GW chirp $\vec{s}_0=\vec{\Pi}
\vec{p_0}$ be \footnote{If the noise power is not unity, it would
  divide the signal energy in this expression. When we have only one
  detector, the SNR $\rho^2$ is consistent with the definition usually
  adopted in this case.}  
\be
\label{eq:rho2}
\rho^2 \equiv \hat{\Lambda}(\vec{s}_0).
\ee

Note that in this expression, the matrix $\vec{\Pi}$ in the statistic
and in $\vec{s}_0$ are the same. Using the SVD of $\vec{\Pi}$ and the
property of the projection operator $\vec{U}_\Pi^H
\vec{U}_\Pi=\vec{I}_r$, we get $\rho^2 = \big\| \vec{s}_0 \big\|^2$.
The SNR is equal to the ``signal energy'' in the network data as
defined in Eq.~(\ref{eq:rho}). Thus, the SNR $\rho$ scales as
$\sqrt{N}$ as expected and it depends on the source direction,
polarization and network configuration through the gain factor $\big
\| \vec{\tilde{f}} \big \|$ \footnote{The SNR $\rho^2$ is similar to
  $b^2$ defined in Eq.~(3.17) of \cite{Pai:2000zt} in case of
  inspiralling binary signal and colored noise.}.
Fig.~{\ref{fig:SNR_net} illustrates how this factor varies for the
  network formed by the two LIGO detectors and Virgo.
  Fig.~{\ref{fig:SNR_net} displays the ratio
    $\rho/\rho_{\mathrm{best}}$ between the global SNR (obtained with
    a coherent analysis) and the largest individual SNR (obtained with
    the best detector of the network). The panels (a) and (b) are
    associated to the ``worst'' (minimum over all polarizations angles
    $\epsilon$ and $\psi$) and ``best'' (maximum) cases respectively.
    Ideally, when the detectors are aligned, the enhancement factor is
    expected to be $\sqrt{d}$ ($\approx 1.73$ in the present case).
    In the best case, the enhancement is $\gtrsim 1.7$ for more than
    half of the sky ($94 \%$ of the sky when $\gtrsim 1.4$).  In the
    worst case, the SNR enhancement is $1.28$ at most and $8.5 \%$ of
    the sky gets a value $ \gtrsim 1.1$.

\begin{figure}
\begin{tabular}{cc}
\includegraphics[width=0.5\columnwidth]{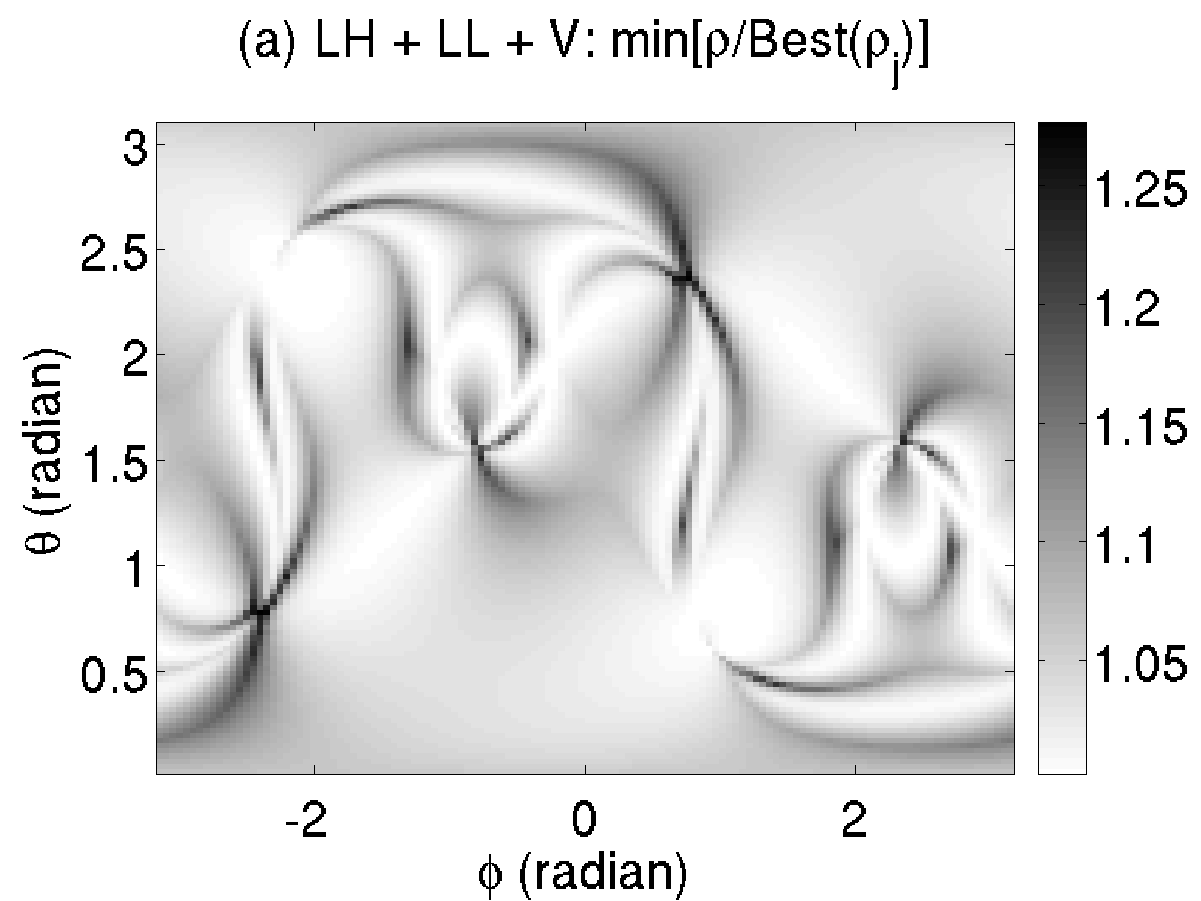}&\includegraphics[width=0.5\columnwidth]{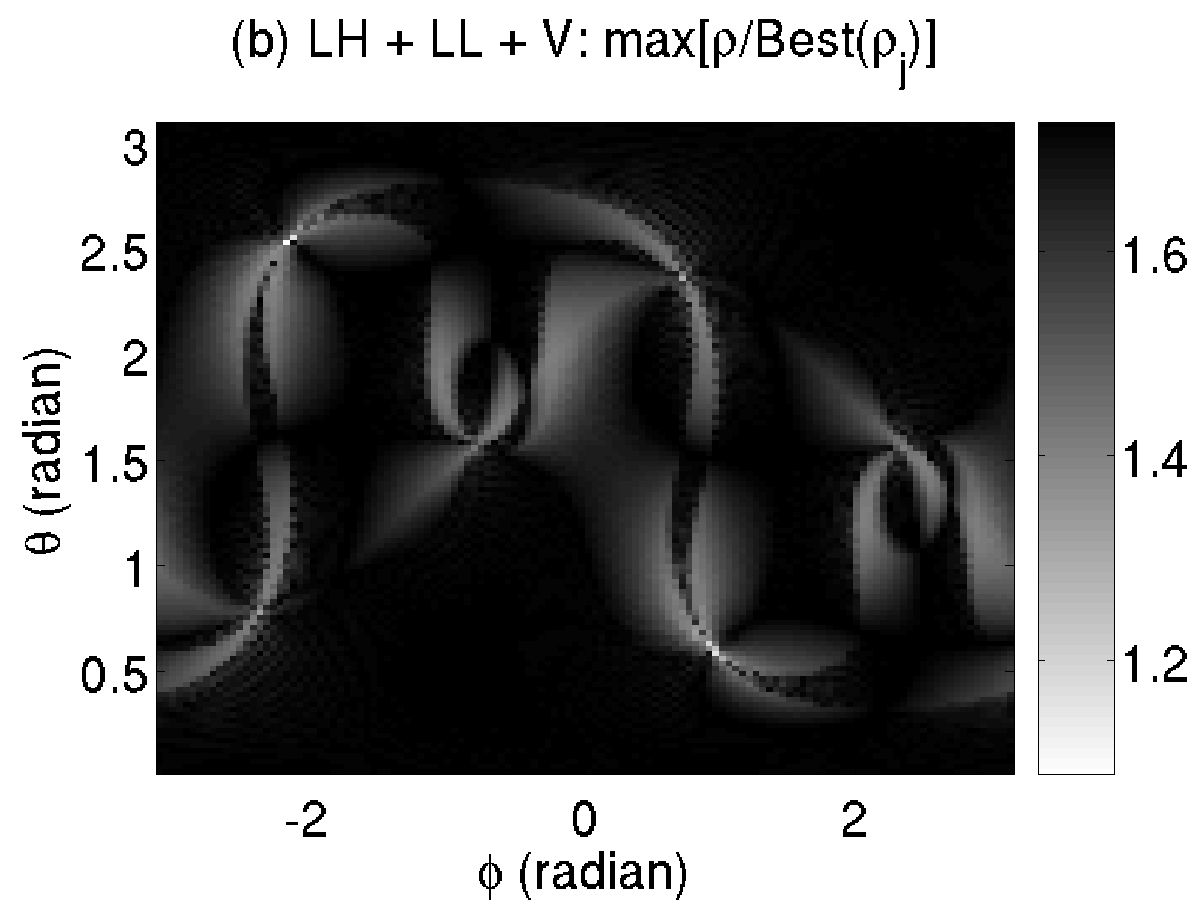}
\end{tabular}
\caption{\textbf{Benefits of a coherent network analysis (SNR
    enhancement).}  We display the polar maps of the following
  quantities for the LL-LH-V network (a) $\min_{\psi,\epsilon}
  \rho/\rho_{\mathrm{best}}$ and (b) $\max_{\psi,\epsilon}
  \rho/\rho_{\mathrm{best}}$.  Here, $\rho_{\mathrm{best}}$ designates
  the best SNR of the detectors in the network.  The maximum, minimum
  are taken over all the polarization angles $\{\psi,\epsilon\}$.}
\label{fig:SNR_net} 
\end{figure}

\subsubsection{From geometrical to physical parameter estimates}
\label{sec:est_extrinsic}
The components of $\vec{p}$ do not have a direct physical
interpretation but as mentioned earlier, they are rather functions of the
physical parameters. Following the above discussion,
if we assume that we obtained parameter estimates $\hat{\vec{p}}$ from the
data through Eq.~(\ref{eq:lsq}), then one can retrieve the physical 
parameters $A$, $\phi_0$, $\epsilon$ and $\psi$ by inverting the
non-linear map which links $\vec{p} = \left(a~t^*_-~~ a^*t_+~~a~t^*_+~~a^*t_-\right)^T$ to these parameters as
given below
\bea
\hat{A} &=& (\sqrt{|\hat{p}[1]|} + \sqrt{|\hat{p}[2]|})^2 \,,\\
\hat{\phi}_0 &=& \frac{1}{2} \left[\arg(\hat{p}[1]) - \arg(\hat{p}[2]) \right] \,,\\
\hat{\psi} &=& -\frac{1}{4} \left[\arg(\hat{p}[1]) + \arg(\hat{p}[2]) \right] \,,\\
\hat{\epsilon} &=& \cos^{-1} \left[ \frac{\sqrt{|\hat{p}[2]|} - \sqrt{|\hat{p}[1]|}}
{\sqrt{|\hat{p}[2]|} + \sqrt{|\hat{p}[1]|}} \right] \,.
\eea

\subsubsection{Degeneracy and sensitivity of the estimate to noise}
Upper-bounds for the estimation error can be obtained using a
perturbative analysis of the LS problem in Eq.~(\ref{eq:mlike}). A
direct use of the result of \cite{golub96:_matrix}, Sec. 5.3.8 yields
\begin{equation}
\frac{\| \hat{\vec{p}}-\vec{p} \|}{\| \vec{p} \|} \leq \frac{\sqrt{N}}{\rho} \cond(\vec{\Pi}).
\end{equation}

This bound is a worst-case estimate obtained when the noise term which
affects the data $\vec{x}$ is essentially concentrated along the
directions associated to the smallest SV of $\vec{\Pi}$.  The noise is
random and it spans isotropically all $Nd$ dimensions of the signal
space. As described above, the space associated to the smallest SV has
only 2 dimensions. Therefore, the worst case is very unlikely to occur
and the above bound is largely over-estimated on the average. However,
it gives a general trend and shows that the estimation goes worst
with the conditioning of $\vec{\Pi}$.

Regularization techniques seem to give promising results in the
context of GW burst detection
\cite{klimenko05:_const,mohanty06:_variab,rakhmanov06:_rank_tikhon}.
Following this idea, we may consider to ``regularize'' the LS problem
in Eq.~(\ref{eq:mlike}). To do so, additional information on the
expected parameters is required to counterbalance the rank-deficiency.
Unfortunately, we don't expect $\vec{p}$ to follow a specific
structure. The only sensible prior that can be assumed without
reducing the generality of the search is that $\| \vec{p} \|$ is
likely to be bound (since the GW have a limited amplitude $A$). It is
known \cite{neumaier98:_solvin_ill} that this type of prior is
associated to the use of the so-called Tykhonov regulator and that we
don't expect significant improvements upon the non-regularized
solution.

Probably, one difference might explain why regularization techniques
do not work in the present case while it does work for burst
detection. We recall that in the burst case, the parameter vector
comparable to $\vec{p}$ are the samples of the waveform.  This vector
being a time-series, it is expected to have some structure, in
particular it is expected to have some degree of smoothness. The use
of this \textit{a priori} information improves significantly the final
estimation.

While regularization will not help for the \textit{estimation} of the
extrinsic parameters, they may be of use to improve the
\textit{detection} statistic. We consider this separate question later
in Sec. \ref{sec:variants}.


\subsection{Implementation with synthetic streams}
\label{sec:synthetic}

In the previous section, we maximize the network LLR with respect to
the extrinsic parameters resulting in the statistic $\hat{\Lambda}$ in
Eq.~(\ref{eq:lam}).
Here, we obtain a more simple and practical expression for
$\hat{\Lambda}$ which will be useful for maximization over the
remaining intrinsic parameters.

From Eqs.~(\ref{eq:svdpi}) and (\ref{eq:lam}), we have
\be
\hat{\Lambda}(\vec{x})=\big\| (\vec{U}_D \otimes \vec{U}_E)^H \vec{x} \big\|^2.
\label{eq:lam_kron}
\ee

It is useful to reshape the network data $\vec{x}$ into a $N \times d$
matrix $\vec{X}\equiv[\vec{x}_1 \vec{x}_2 \ldots \vec{x}_d]$.
This operation is inverse to the \textit{stack operator} $\stack{}$ defined in
Appendix \ref{app:kron}.

Using a property of the Kronecker product in Eq.
(\ref{eq:kron_stack}), we obtain the reformulation
\be
\hat{\Lambda}(\vec{x})=\big\| \stack{\vec{U}_E^H \vec{X} \vec{U}_D^*} \big\|^2.
\ee

There are two possibilities to make this matrix product, each being
associated to a different numerical implementation for the evaluation
of $\hat{\Lambda}$.

A first choice is to first multiply $\vec{X}$ by $\vec{U}_E^H$ and then by
$\vec{U}_D^*$. In practice, this means that we first compute the
correlation of the data with a chirp template, then we combine the
result using weights (related to the antenna pattern functions). This
is the implementation proposed in \cite{Pai:2000zt}. It is probably
the best for cases (like, searches of inspiralling binary chirps) where
the number of chirp templates is large (i.e., larger than the number
of source locations) and where the correlations with templates are
computed once and stored.

The second choice is to first multiply $\vec{X}$ by $\vec{U}_D^*$ and then by
$\vec{U}_E^H$ which we adopt here.
 This means that we first compute $\vec{Y} \equiv
\vec{X}\vec{U}_D^*$ which transforms the network data into two
$N$-dimensional complex data vectors $[\vec{y}_1,
\vec{y}_2] \equiv \vec{Y}$ through an ``instantaneous'' linear
combination. Then, we correlate these vectors with the chirp template.
We can consider $\vec{y}_1$ and $\vec{y}_2$ as the output of two
``virtual'' detectors. For this reason, we refer to those as
\textit{synthetic streams} in connection to \cite{sylvestre03} who
first coined the term for such combinations.  Note that, irrespective
of the number of detectors, one always gets at most two synthetic
streams.  We note that though the synthetic streams defined in
\cite{sylvestre03} are {\it ad-hoc} (i.e., they have no relation with
the LR), the ones obtained here directly arise from the maximization
of the network LLR.

We express the network LLR statistic in terms of the two
synthetic streams as 
\be
\hat{\Lambda}(\vec{x})=\frac{1}{N}\left(|\vec{e}^H \vec{y}_1|^2 + |\vec{e}^T \vec{y}_1|^2 + |\vec{e}^H \vec{y}_2|^2 + |\vec{e}^T \vec{y}_2|^2 \right),
\ee
where $\vec{y}_{l}=\vec{X}\vec{u}^*_{l}$ for $l=1,2$. This expression can be further 
simplified by using the symmetry (easily seen
from Eqs. (\ref{eq:u1}) and (\ref{eq:u2})),
\be
\vec{u}_1^* = \exp(i \delta) \vec{u}_1 \,, \quad \quad \vec{u}_2^* = -\exp(i \delta) \vec{u}_2 \,.\ee

We finally obtain
\be
\label{eq:netMLR}
\hat{\Lambda}(\vec{x}) = \frac{2}{N}(|\vec{e}^H \vec{y}_1 |^2 +| \vec{e}^H \vec{y}_2 |^2) \,.
\ee

The linear combination in each stream is such that the signal
contributions from each detector add up constructively.  In this
sense, synthetic streams are similar to beam-formers used in array
signal processing \cite{van02:_optim}.

When the data is a noise free GW chirp, i.e., $\vec{x}=\vec{s}$, we
then have 
\be
y_l[k] = \vec{p}^T (\vec{D} \otimes \vec{E}[k])^T \vec{u}^*_l = \vec{p}^T (\sigma_l \vec{v}^*_l \otimes \vec{E}[k]^T)\,. 
\label{eq:synth_noisefree}
\ee

Here, $\vec{E}[k]$ represents the $k$-th row of $\vec{E}$. Writing explicitly, we have
\begin{align}
\vec{y}_1 &=\frac{\sigma_1}{\sqrt{2}} \Re\{q_1 \vec{\tilde{h}} \} e^{i\delta/2},&
\vec{y}_2 &= \frac{i \sigma_2}{\sqrt{2}}\Im\{q_2 \vec{\tilde{h}} \} e^{i\delta/2} \label{eq:synth_noisefree_bis} \,,
\end{align}
where $q_{1}=(t_- e^{-i\delta/2} + t_+ e^{i\delta/2})^*$, $q_{2}=(t_-
e^{-i\delta/2} - t_+ e^{i\delta/2})^*$ and where $\vec{\tilde{h}}$ is
as defined in Eq.~(\ref{eq:netS2}). This shows that the synthetic
streams $\vec{y}_l$ are rescaled and phase shifted copies of the
initial GW chirp $\vec{\tilde{h}}$ which enhances the amplitude of the
signal by the appropriate factor as shown above.

To assess this enhancement, we compute the SNR per synthetic stream as
below.

\subsubsection{SNR per synthetic streams}
\label{sec:snr_per_synth_streams}

The network SNR can be split into the contributions from each synthetic stream
{\it i.e.} we write $\rho^2=\big\|\vec{\Sigma}_\Pi \vec{V}_\Pi^H \vec{p}_0 \big\|^2$, 
as
\be
\rho^2 = \rho_1^2 + \rho_2^2,
\ee 
where we define $\rho_l \equiv \sqrt{N} \sigma_l \big\| (\vec{v}_l \otimes
\vec{I}_2)^H \vec{p}_0 \big\|/2$ for $l=1,2$. 
More explicitly, we have
\be
\rho_{l} = \frac{\sqrt{N}}{2} \sigma_{l} |q_{l}| A\,.
\label{eq:snr_per_synth}
\ee

\begin{figure}
\begin{tabular}{cc}
\includegraphics[width=0.5\columnwidth]{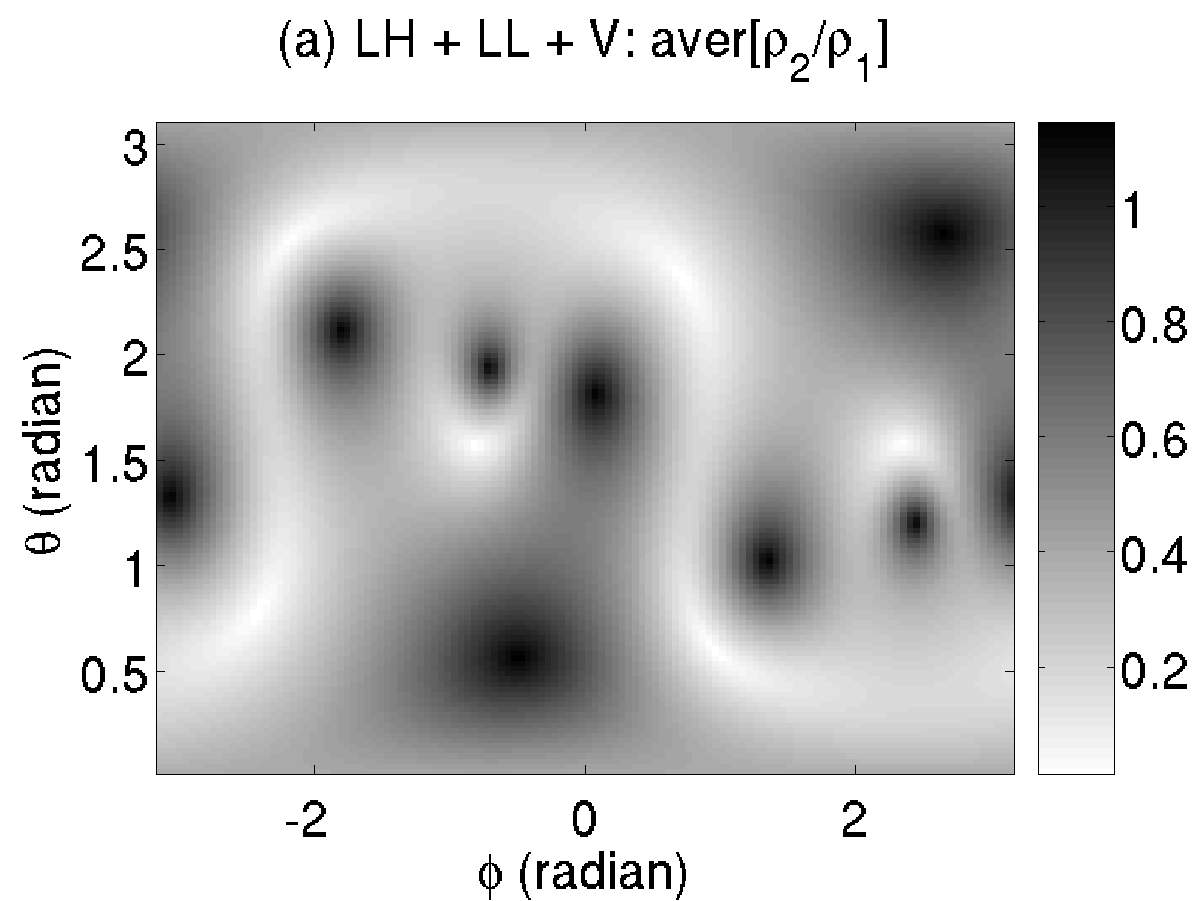}&
\includegraphics[width=0.5\columnwidth]{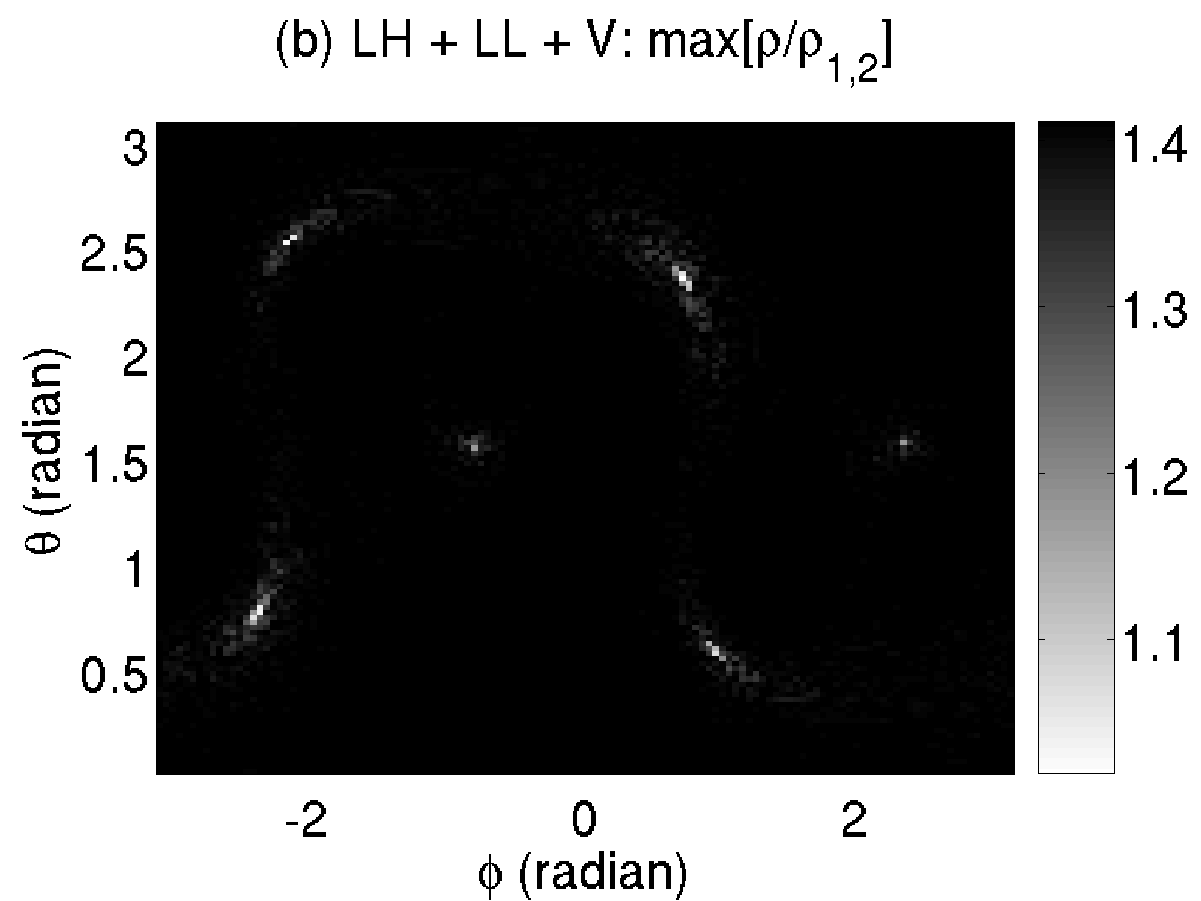}
\end{tabular}
\caption{\label{fig:SNRratio_net} \textbf{Color online --- SNR per
    synthetic streams and benefits of a coherent network analysis (SNR
    enhancement).}  We display the polar maps of the following
  quantities for the LL-LH-V network: (a)
  $\lag\rho_2/\rho_1\rag_{\psi,\epsilon}$ and (b)
  $\max_{\psi,\epsilon} \rho/\max_l{\rho_l} $ . We denote $\rho_l$ the
  SNR of synthetic stream as defined in
  Eq.~(\ref{eq:snr_per_synth}).  The maximum and average
  are taken over all the polarization angles $\{\psi,\epsilon\}$.}
\end{figure}

The synthetic streams contributes differently depending on the
polarization of the incoming wave. Fig.~\ref{fig:SNRratio_net}
illustrates this with the network formed by the two LIGO detectors and
Virgo. 

Let us assume that $\vec{p}_0$ is randomly oriented.  Since
$\vec{v}_1$ and $\vec{v}_2$ have unit norms, we get the average value
$0 < \langle \rho_2/\rho_1 \rangle_{\e,\psi} \propto
1/\cond(\vec{\Pi})\le 1$ for most of the sky as indicated in
Fig.~\ref{fig:SNRratio_net} (a). Note that this panel matches well
with Fig.~\ref{fig:cond_network} (b). Thus, on average, $\vec{y}_1$
contributes more to the SNR than $\vec{y}_2$.  However, the situation
may be different depending on the specific polarization state of the
wave. Fig.~{\ref{fig:SNRratio_net} (b) show the maximum of the ratio
  $\rho/\max_{l=1,2}{\rho_l}$ for all polarization angles $\epsilon$
  and $\psi$ mostly takes the value $\approx \sqrt{2}$.  This implies
  that for a given direction, one can always find polarization angles
  where the two synthetic streams contribute equally. This holds true
  for all the sky locations, except at the degenerate sky locations
  where $\vec{y}_2$ does not contribute, hence the SNR ratio is $1$.

Another way to see this is to examine the expression of the SNR difference
in terms of the signal and network quantities, namely
\begin{multline}
\rho_l^2 -\rho_2^2 =  \frac{N A^2} {4} \left( \frac{|\Delta|}{2} \left[\left(\frac{1+\cos^2 \e}{2}\right)^2 + \cos^2 \e \right] \right.\\
\left. + \frac{\big \| \vec{f}_+ \big \|^4 - \big \| \vec{f}_\times \big \|^4}{2 |\Delta|} \frac{\sin^4 \e}{4} \right),
\end{multline}
where we have\footnote{The synthetic streams (on the average sense)
  are also connected to the directional streams introduced in the
  context of LISA \cite{Nayak:2003na}. If we integrate $\rho_l^2$ over
  the inclination and polarization angles $\e ,\psi$, we obtain
  $\langle(\rho_1^2-\rho_2^2)/2 \rangle_{\e,\psi} = 2|\Delta|/5$ and
  $\langle \big \| \vec{\tilde{f}} \big \|^2 \rangle_{\e,\psi} = 2
  {\cal D}/5$. Thus, the SNRs of the synthetic streams $\rho_l^2$ when
  averaged over the polarization angles are proportional to the SNRs
  obtained by $v_+$ and $v_\times$ -- the directional streams in the
  LISA data analysis, see Eqs.(25-28) of \cite{Nayak:2003na}.}
$|\Delta |^2= (\big \| \vec{f}_+ \big \|^2 + \big \| \vec{f}_\times
\big \|^2)^2 - 4 \big \| \vec{f}_+ \times \vec{f}_\times \big \|^2$.

In the simple case of a face-on source, {\it i.e.} $\epsilon=0$, and
if $ \vec{f}_\times$ is orthogonal to $\vec{f}_+$ with $ \big \|
\vec{f}_\times \big \| = \big \| \vec{f}_+ \big \|$, then both
synthetic streams carry the same SNR $\rho_1=\rho_2=\sqrt{N/2} A \big
\| \vec{f}_+ \big \|$. However, for other situations, some other wave
polarization would lead the same.

Inversely looking at $\min_{\epsilon,\psi} \rho/\max_l \rho_{l}$, it
can be shown that one can always find a GW polarization such that
one of the synthetic streams does not contribute to the SNR.

\subsection{Null streams}

\subsubsection{Review and relation to synthetic streams}
The access to \textit{noise only} data is crucial in signal detection
problem. Such data is not directly available in GW experiments, but
the use of multiple detectors allows to access it indirectly using the
\textit{null streams}. The general idea behind the null stream is to
construct a data stream from the individual detector streams which
nullifies the signature of any incoming GW from a particular
direction.  Since this signal cancellation is specific to GWs, null
streams naturally provide an extra tool to verify that a detected
signal is indeed a GW or instead a GW like features mimicked by the
detector noise whose detection thus has to be vetoed. This is a
powerful check since it does not require detailed information about
the potential GW signal under test, except an estimate of its source
location. (Note that in practice, the implementation of the veto test
may be complicated by the imprecision of the direction of arrival and
of the errors of calibration \cite{chatterji06:_coher}). The existence
of null streams has been first identified in \cite{Guersel:1990bn} in
the case of three detector networks. At present, handful of literature
\cite{Wen:2005ui,chatterji06:_coher} exists on the use of null streams
in GW data analysis.

Null streams are usually introduced as a general post-processing of
the data independent of the detection of specific GWs. Below, we make
this connection in the domain of our formalism.  We recall that the
network data at a given time (e.g., the first row of the matrix
$\vec{X}$ introduced in Sec. \ref{sec:synthetic}) is a $d$-dimensional
vector in $\R^d$.  This space is a direct sum of the GW polarization
plane and its orthogonal complementary space. We have shown that the
GW polarization plane is a $2$-dimensional space, spanned by a pair of
orthonormal basis vectors which are associated to the two synthetic
streams. The complementary space to the GW polarization plane is a
$d-2$ dimensional space and it is spanned by $d-2$ ``null vectors''.
Similarly to the synthetic streams, the null streams can be
constructed from these null vectors. Thus, the numbers of synthetic
and null streams sum up to $d$. Nominally, we have $d-2$ null streams.
However, when the GW polarization plane degenerates to a
$1$-dimensional space ($\sigma_2 \sim 0$) as explained in
Sec.~\ref{sec:rank_degen}, the number of null vectors becomes $d-1$.
For a two detector network, in the nominal case, there is no null
stream as $d-2 = 0$. However, for degenerate directions, one can
construct a null stream.  For aligned pair of detectors (as it is
almost the case for the two LIGO LH and LL), the fraction of the
degenerate sky location is large, see Fig.~\ref{fig:cond_network}.
This null stream would turn out to be useful for vetoing in this case.

In the next section, we explain how the null streams can be obtained numerically in the
nominal case. The extension to the degenerate case is straightforward.


\subsubsection{Obtaining the null streams numerically}
The numerical construction of the null streams can be achieved in various ways.
One such approach could be to obtain the full SVD of $\vec{D}$ and construct the
null streams from the eigen-vectors corresponding to the zero SVs. 
This approach was taken in \cite{chatterji06:_coher}.
Here, we take an alternative approach. We construct the null streams by successive
construction of orthonormal vectors {\it via} multi-dimensional cross product as described
below.

Assuming some direction of arrival, we express any instantaneous
linear combination of the time-shifted data (to compensate for different
time of arrivals at the detectors' site with respect to the reference) as
\be
\vec{y}(\vec{x}) \equiv \vec{X} \vec{u},
\label{eq:nullstr}
\ee
where the vector $\vec{u}\in \C^{d\times 1}$ contains the tap coefficients.

Eq.~(\ref{eq:nullstr}) can be rewritten as
\be
\vec{y}(\vec{x})=\stack{\vec{I}_N \vec{X} \vec{u}} =(\vec{u}^T \otimes \vec{I}_N) \vec{x}.
\ee

The vector $\vec{u}$ defines a null-stream if
$\vec{y}(\vec{x})=\vec{0}_N$ whenever $\vec{x}$ is a GW. Let us assume
that we indeed observe a GW chirp i.e.,
$\vec{x}=\vec{s}_0\equiv\vec{\Pi}\vec{p}_0$. We thus have
\be
\vec{y}(\vec{s}_0)= [(\vec{u}^T \vec{U}_D) \otimes \vec{U}_E] \vec{\Sigma}_{\Pi} \vec{V}_{\Pi}^H \vec{p}_0.
\ee

If $\vec{u}$ is in the null space of $\vec{U}_D$, the
null-stream condition is satisfied for all $\vec{p}_0$. Since the null
space of $\vec{U}_D$ is orthogonal to its range, an obvious choice for
$\vec{u}$ is
\be
\vec{u} = \vec{u}_1 \times \vec{u}_2 = \frac{\vec{d}^* \times \vec{d}}{\sqrt{{\cal D}^2 - \Delta^2}}.
\ee
Nominally, $\vec{U}_D$ is a 2-dimensional plane in $\C^d$.
Its null space is therefore $d-2$ dimensional. An orthonormal
basis of this space can be obtained recursively starting from
$\vec{u}_3=\vec{u}$ as defined above and applying the following
generalized vector cross-product formula for $n>3$:
\be
\vec{u}_n[i] = \e_{ijkl\ldots m} \vec{u}_1[j] \vec{u}_2[k] \vec{u}_3[l] \ldots \vec{u}_{n-1}[m].
\ee

Here, $\e_{ijkl\ldots m} $ is the Levi-Civita symbol
\footnote{The Levi-Civita symbol is defined as
\begin{align*}
\e_{ij\ldots} &\equiv +1 &&\text{when $i, j,\ldots$ is an \textit{even} permutation of $1, 2, \ldots$}\\
&\equiv -1  &&\text{when $i, j, \ldots$ is an \textit{odd} permutation of $1, 2, \ldots$}\\
&\equiv 0   &&\text{when any two labels are equal.}
\end{align*}}.
The $\vec{u}_n$ denotes an orthonormal set of $d-2$ vectors, $\{\vec{u}_{n},
\text{~for~~} 3\leq n\leq d\}$. The components of these vectors are
the tap coefficients to compute the null-streams. By construction, the
resulting null streams are uncorrelated and have the same variance.

To summarize the main features of our formalism. The representation of 
a GW network response of unmodeled chirp as a Kronecker product between 
the GW polarization plane and the chirp plane forms the main ingredient 
of this formalism. Such a representation allows the signal to reveal
the degeneracy in a natural manner in the network response. It also
evidences the two facets of the coherent network detection problem,
namely, the network signal detection {\it via} synthetic streams
and vetoing {\it via} null streams. The coherent formalism developed
in \cite{Pai:2000zt} for inspiralling binaries lacked this vetoing feature due to
the difference in the signal representation. 

In the rest of this paper, we do not discuss/demonstrate
the null streams applied as a vetoing tool
to the simulated data. This will be demonstrated in the subsequent 
work with the real data from the ongoing GW experiments.

\subsection{Colored noise}
\label{sec:color}  
The formalism developed till now was exclusively targeted for the white
noise case. We assumed that the noise at each detector is white Gaussian.
In this subsection, we extend our formalism to the colored
noise case. We remind the reader that the main focus of this paper is
to develop the coherent network strategy to detect unmodeled GW chirps
with an interferometric detector network. Hence, we give more emphasis on
the basic formalism and keep the colored noise case with basic minimal
assumption: the noise from the different detectors is colored but with same 
covariance. Based on this ground work, the work is in progress to extend
this to the colored noise case with different noise covariances.

Let us therefore assume now that the noise components in each detector are independent and colored, 
with the same covariance matrix $\vec{R}_0$. Recall that the 
 covariance matrix of a random vector $\vec{a}$ is defined as $\mathbb{E} \left[ (\vec{a}-\mathbb{E}[\vec{a}]) (\vec{a}-\mathbb{E}[\vec{a}])^H \right]$ where $\mathbb{E}[.]$ denotes the expectation. From the independence of the noise components, the overall covariance matrix of the network noise vector is then a block-diagonal matrix, where all the blocks are identical and equal to $\vec{R}_0$: $\vec{R} = \diag (\vec{R}_0) \equiv \vec{I}_d \otimes \vec{R}_0$.

 In this case, the network LLR becomes
\be
\Lambda(\vec{x}) =-\big\| \vec{x} - \vec{\Pi} \vec{p} \big\|_{\vec{R}^{-1}}^2 + \big\| \vec{x} \big\|_{\vec{R}^{-1}}^2, \label{eq:mlr_colored}
\ee  
where the notation $\big\|\cdot\big\|_{\vec{R}^{-1}}^2$ denotes the
norm induced by the inner product associated to the covariance matrix ${\vec{R}^{-1}}$, i.e., $\big\| \vec{a} \big\|_{\vec{R}^{-1}}^2 = \vec{a}^H \vec{R}^{-1} \vec{a}$.

Introducing the whitened version
$\tilde{\vec{\Pi}}=\vec{R}^{-1/2}\vec{\Pi}$ and
$\tilde{\vec{x}}=\vec{R}^{-1/2}\vec{x}$ of $\vec{\Pi}$ and $\vec{x}$
respectively, Eq.~(\ref{eq:mlr_colored}) can be rewritten as
\be
\Lambda(\vec{x}) =-\big\| \tilde{\vec{x}} - \tilde{\vec{\Pi}}\vec{p} \big\|^2 + \big\| \tilde{\vec{x}} \big\|^2. \label{eq:mlr_colored_whitened}
\ee  
which is similar to Eq.~(\ref{eq:mlr}) where all the quantities are replaced
by their whitened version. Thus, the maximization of $\Lambda(\vec{x})$
with respect to the extrinsic parameters $\vec{p}$ will follow the same algebra
as that derived in \ref{sec:LR}. However, for the sake of completeness, we detail it below.

Following Sec.~\ref{sec:LR}, maximizing $\Lambda(\vec{x})$
with respect to the extrinsic parameters $\vec{p}$ leads to
\be
\label{eq:lsq_colored}
\hat{\vec{p}} = \tilde{\vec{\Pi}}^\# \tilde{\vec{x}},
\ee
where $\tilde{\vec{\Pi}}^\#$ is the pseudo-inverse of
$\tilde{\vec{\Pi}}$. Expressing this pseudo-inverse by means of the
SVD of $\tilde{\vec{\Pi}}$ as
$\tilde{\vec{\Pi}}^\#=\vec{V}_{\tilde{\Pi}}
\vec{\Sigma}_{\tilde{\Pi}}^{-1} \vec{U}_{\tilde{\Pi}}^H$ and
introducing Eq.~(\ref{eq:lsq_colored}) into
Eq.~(\ref{eq:mlr_colored_whitened}) provides the new statistic
\be
\label{eq:lam_colored}
\hat{\Lambda}(\vec{x}) = \big\| \vec{U}_{\tilde{\Pi}}^H \tilde{\vec{x}} \big\|^2.
\ee

Now, from the definition of $\vec{\Pi}$ and the specific structure of
$\vec{R}$, it is straightforward to see that
\be
\tilde{\vec{\Pi}} = \vec{R}^{-1/2} (\vec{D} \otimes \vec{E}) = \vec{D} \otimes (\vec{R}_0^{-1/2}\vec{E}) = \vec{D} \otimes \tilde{\vec{E}},
\ee
where we have introduced the whitened version $\tilde{\vec{e}} =
\vec{R}_0^{-1/2} \vec{e}$ of the chirp signal and the corresponding
matrix $\tilde{\vec{E}} = [\tilde{\vec{e}}~ \tilde{\vec{e}}^*]/2$.

For the white noise case, the statistic (\ref{eq:lam_colored}) can
then be rewritten in terms of the SVD of the matrices $\vec{D}$ and
$\tilde{\vec{E}}$:
\be
\hat{\Lambda}(\vec{x})=\big\| (\vec{U}_D \otimes \vec{U}_{\tilde{E}})^H \tilde{\vec{x}} \big\|^2.
\label{eq:lam_kron_colored}
\ee 

The computation of $\vec{U}_{\tilde{E}}$ is similar to the computation
of $\vec{U}_{\tilde{E}}$. Furthermore, if we note that
$\tilde{\vec{e}}^T \tilde{\vec{e}} \simeq 0$, and if we assume that
$\tilde{\vec{e}}^H \tilde{\vec{e}} = N$, it turns out that
$\vec{U}_{\tilde{E}} = 2 \tilde{\vec{E}} / \sqrt{N} =
2 \vec{R}_0^{-1/2}\vec{E} / \sqrt{N}$. Using the property of the
Kronecker product in Eq.~(\ref{eq:kron_stack}), we then obtain
\be
\hat{\Lambda}(\vec{x})=\big\| \stack{\vec{E}^H \tilde{\tilde{\vec{X}}} \vec{U}_D^*} \big\|^2,
\ee
where the matrix $\tilde{\tilde{\vec{X}}} = [\tilde{\tilde{\vec{x}}}_1
\tilde{\tilde{\vec{x}}}_2 \dots \tilde{\tilde{\vec{x}}}_d]$ contains
the data vector from each detector whitened twice:
$\tilde{\tilde{\vec{x}}}_j = \vec{R}^{-1/2}\tilde{\vec{x}}_j =
\vec{R}^{-1}\vec{x}_j$.

As this expression is similar to the white noise case, we can form two
synthetic streams $[\tilde{\tilde{\vec{y}}}_1,\tilde{\tilde{\vec{y}}}_2] =
\tilde{\tilde{\vec{X}}} \vec{U}_D^*$ and use them to express the LLR
statistic as 
\be 
\hat{\Lambda}(\vec{x}) = \frac{2}{N}(|\vec{e}^H
\tilde{\tilde{\vec{y}}}_1 |^2 +| \vec{e}^H \tilde{\tilde{\vec{y}}}_2 |^2).
\label{eq:netMLR_colored}
\ee

In this expression, the only difference with the white noise LLR of
(\ref{eq:netMLR}) comes from the computation of the synthetic streams
$\tilde{\tilde{\vec{y}}}_1$ and $\tilde{\tilde{\vec{y}}}_2$ which are
obtained after double-whitening the data.

\section{Maximization over the intrinsic parameters}
\label{sec:intrinsic}

In the previous section, we maximized the network LLR over the
extrinsic parameters of the signal model, assuming that the remaining
parameters (the source location angles $\phi$ and $\theta$ and the
phase function $\varphi(\cdot)$) were known.

By definition, the intrinsic parameters modify the network LLR
non-linearly. For this reason, the maximization of $\hat{\Lambda}$
over these parameters is more difficult. It cannot be done
analytically and must be performed numerically, for instance with an
exhaustive search of the maximum by repeatedly computing
$\hat{\Lambda}$ over the entire range of possibilities.

While the exhaustive search can be employed for the source location
angles, it is not applicable to the chirp phase function, which
requires a specific method. For the single detector case, we had
addressed this issue in \cite{chassande-mottin06:_chirplet_chains}
with an original maximization scheme which is the cornerstone of the
best chirplet chain (BCC) algorithm.  Here, we use and adapt the
principles of BCC to the multiple detector case.

\subsection{Chirp phase function}


Let us examine first the case of the detection of inspiralling
binary chirps. In this case, the chirp phase is a prescribed
function of a small number of parameters i.e., the masses and spins of
the binary stars. The maximization over those is performed by
constructing a grid of reference or \textit{template} waveforms which
are used to search the data.  This grid samples the range of the
physical parameters. This sampling must be accurate (the template grid
must be tight) to avoid missing any chirp.

Tight grid of templates can be obtained in the non-parametric case
(large number of parameters) i.e., when the chirp is not completely
known.  We have shown in \cite{chassande-mottin06:_chirplet_chains} how to
construct a template grid which covers entirely the set of smooth
chirps i.e., chirps whose frequency evolution has some regularity as
described in Sec.~\ref{sec:smooth_chirps}. In the next section, we
briefly describe this construction.

\subsubsection{Chirplet chains: tight template bank for smooth chirps} 
We refer to the template forming this grid as \textit{chirplet
  chain} (CC). These CCs are constructed on a simple geometrical idea: a broken
line is a good approximation of a smooth curve. Since the frequency of
a smooth chirp follows a smooth frequency \textit{vs.} time curve, we
construct templates that are broken lines in the time-frequency (TF) plane.


More precisely, CC are defined as follows. We start by sampling the
TF plane with a regular grid consisting of $N_t$ time bins
and $N_f$ frequency bins. We build the template waveforms like a
puzzle by assembling small chirp pieces which we refer to as
\textit{chirplets}. A chirplet is a signal with a frequency joining
linearly two neighboring vertices of the grid. The result of this
assembly is a \textit{chirplet chain} i.e., a piecewise linear
chirp. Since we are concerned with continuous frequency evolution with
bounded variations, we only form continuous chains.

We control the variations of the CC frequency. The frequency of a
single chirplet does not increase or decrease more than $N_r'$ frequency
bins over a time bin. Similarly, the difference of the frequency
variations of two successive chirplets in a chain does not increase or
decrease more than $N_r''$ frequency bins.

The CC grid is defined by four parameters namely $N_t$, $N_f$, $N'_r$
and $N''_r$. Those are the available degrees of freedom we can tune to
make the CC grid tight. A template grid is tight if the network ambiguity
$\hat{\Lambda}(\vec{s};\varphi') \propto |{\vec{e}'}^H
\vec{y}_1(\vec{s})|^2 +|{\vec{e}'}^H \vec{y}_2(\vec{s})|^2$ which
measures the similarity between an arbitrary chirp (of phase
$\varphi$) and its closest template (of phase $\varphi'$), is large
enough and relatively closed to the maximum (when $\varphi'=\varphi$).

As stated in Eq.~(\ref{eq:synth_noisefree_bis}), in presence of a noise
free GW chirp, the synthetic streams $\vec{y}_l(\vec{s})$ are rescaled and
phase shifted copies of the initial GW chirp $\vec{s}$. Therefore, 
we treat the network ambiguity as a sum of ambiguities
from two {\it virtual} detectors where each terms in $\hat{\Lambda}(\vec{s};\varphi')$
is the ambiguity computed in the single detector case. An estimate of the ambiguity has been
obtained in \cite{chassande-mottin06:_chirplet_chains} for this case.
It can thus be directly reused to compute
$\hat{\Lambda}(\vec{s};\varphi')$.

The bottom line is that the ratio of the ambiguity to its maximum
for the network case remains unaltered as compared to the single detector case and thus
same for the tight grid conditions. In conclusion, the rules (which we
won't repeat here) established in
\cite{chassande-mottin06:_chirplet_chains} to set the search
parameters can also be applied here.

\subsubsection{Search through CCs in the time-frequency plane: best network CC algorithm} 


We have now to search through the CC grid to find the best matching
template, i.e., which maximizes $\max_{\text{all CCs
    $\varphi'$}} \hat{\Lambda}(\vec{x};\varphi')$. Counting the number
of possible CCs to be searched over is a combinatorial problem.
This count grows exponentially with the number of time bins $N_t$.
In the situation of interest, it
reaches prohibitively large values.  The family of CCs cannot be
scanned exhaustively and the template based search is intractable.

In \cite{chassande-mottin06:_chirplet_chains}, we propose an
alternative scheme yielding a close approximation of the maximum
for the single detector statistic. When applied to the network, the
scheme demands to reformulate the network statistic in the
TF plane. The TF plane offers a natural and
geometrically simple representation of chirp signals which simplifies
the statistic. It turns out that the resulting statistic
falls in a class of objective functions where efficient combinatorial
optimization algorithms can be used. We now explain this result in more
detail.

We use the TF representation given by the discrete Wigner-Ville
(WV) distribution \cite{Eric:2005a} defined for the time series $x[n]$
with $n=0,\ldots,N-1$ as 
\be
  \label{dwv}
  w_x(n,m) \equiv \sum_{k=-k_n}^{+k_n} x[\lfloor n+k/2 \rfloor]  x^*[\lfloor n-k/2 \rfloor] \, e^{-2\pi i mk/(2N)},
\ee
with $k_n\equiv \min\{ 2n, 2N-1-2n \}$, 
where $\lfloor
\cdot \rfloor$ gives the integer part. The arguments of $w_x$ are the
time index $n$ and the frequency index $m$ which correspond, in
physical units, to the time $t_n=t_s n$ and the frequency $\nu_m=\nu_s
m/(2N)$ for $0\leq m\leq N$ and $\nu_m=\nu_s (N-m)/(2N)$ for $N+1\leq
m\leq 2N-1$.  

The above WV distribution is a unitary representation. This means that the
scalar products of two signals can be re-expressed as scalar products
of their WV. Let $x_1[n]$ and $x_2[n]$ be two time series. The unitarity
property of  $w_x$ is expressed by the Moyal's formula as stated below
\begin{equation}
\label{moyal3}
\left|\sum_{n=0}^{N-1} x_1[n] \, x_2^*[n] \right|^2=
\frac{1}{2N} \sum_{n=0}^{N-1} \sum_{m=0}^{2N-1} w_{x_1}(n,m)\, w_{x_2}(n,m).
\end{equation}

Applying this property to the network statistic in
Eq.~(\ref{eq:netMLR}), we get
\begin{equation}
\label{moyal}
\hat{\Lambda}(\vec{x})=\frac{1}{N^2}\sum_{n=0}^{N-1}\sum_{m=0}^{2N-1} w_y(n,m) w_e(n,m).
\end{equation}
where $w_y=w_{y_1}+w_{y_2}$ combines the individual WVs of the
two synthetic streams.

In order to compute $\hat{\Lambda}(\vec{x})$, we need to have a model
for $w_e$. We know that the WV distribution of a linear chirp
(whose frequency is a linear function of time) is essentially concentrated
in the neighborhood of its instantaneous frequency \cite{Eric:2005a}.
We assume that it also holds true for an arbitrary (non-linear) chirp.
Applying this approximation to the WV $w_e$ of the template CC in
Eq.~(\ref{moyal}), we get
\begin{equation}
\label{dwv_model}
w_e(n,m)\approx 2N \:\delta(m - m_n),
\end{equation}
where $m_n$ denotes the nearest integer of $2T\:\nu(t_n)$
and $\nu$ is the instantaneous frequency of the CC.

Thus, substituting in Eq.~(\ref{moyal}), we obtain the following
reformulation of the network statistic
\begin{equation}
\label{eq:pathint}
\hat{\Lambda}(\vec{x})= \frac{2}{N} \sum_{n=0}^{N-1} w_y(n,m_n).
\end{equation}

The maximization of $\hat{\Lambda}(\vec{x})$ over the set of CC
amounts to finding the TF path that maximizes the integral
Eq.~(\ref{eq:pathint}), which is
equivalent to a \textit{longest path problem} in the TF plane.  
This problem is structurally identical to the
single detector case (the only change is the way we obtain the TF
map). We can therefore essentially re-use the scheme proposed earlier
for this latter case.  The latter belongs to
a class of combinatorial optimization problems where efficient
(polynomial time) algorithms exist. We use one such algorithm,
namely, the dynamic programming.

In conclusion, \textit{the combination of the two ingredients namely
the synthetic streams and the phase maximization scheme used in BCC allows
us to coherently search the unmodeled GW chirps in the data of GW
detector network.} We refer to this procedure as the \textit{best network CC} (BNCC) algorithm.

\subsection{Source sky position} 

As we are performing maximization successively, till now we assume
that we know the sky position of the source. Knowing the sky position,
we construct the synthetic streams with appropriate direction
dependent weight factors, time-delay shifts and carry out the BNCC
algorithm for chirp phase detection. In reality, the sky
position is unknown. One needs to search through the entire sky by
sampling the celestial sphere with a grid and repeating the above
procedure for each point on this grid.

\subsection{Time of arrival} 

Since we process the data streams sequentially and block-wise, the
maximization over $t_0$ amounts to selecting that block where the
statistic arrives at a local maximum (i.e., the maximum of the
``detection peak'').  The epoch of this block yields an estimate of
$t_0$. The resolution of the estimate may be improved by increasing
the overlap between two consecutive blocks.

\subsection{Estimation of computational cost}

We estimate the computational cost of the BNCC search by counting the
floating-point operations (flops) required by its various subparts.
The algorithm consists essentially in repeating the one-detector
search for all sky location angles. Let $N_\Omega$ be the number of
bins of the sky grid. The total cost is therefore $N_\Omega$ times the
cost of the one-detector search, which we give in
\cite{chassande-mottin06:_chirplet_chains} and summarize now.  The
computation of the WV of the two synthetic streams requires $10 N N_f
\log_2 N_f$ flops and the BCC search applied to the combined WV
requires $[N+(2N_r''+1)N_t]N_c$ flops, where $N_c\approx (2N_r'+1)N_f$
is the total number of chirplets. Since this last part of the algorithm
dominates, the over-all cost thus scales with
\be
\label{eq:cost}
C \propto N_\Omega [N+(2N_r''+1)N_t] (2N_r'+1) N_f\,.
\ee

These operations are applied to each blocks (of duration $T$) the data
streams. The computational power needed to process the data in real
time is thus given by $C/(\mu T)$ where $\mu$ is the overlap between
two successive blocks.

\section{Results with simulated data and discussion}
\label{sec:discuss}

\subsection{Proof of principle of a full blind search}
We present here a proof of principle for the proposed detection method.
For this case study, we consider a network of three detectors placed
and oriented like the existing Virgo and the two four-kilometer LIGO
detectors. The coordinates and orientation of these detectors can be
found in Table \ref{tab:det_euler}. We assume a simplified model for
detector noise which we generate independently for each detector,
using a white Gaussian noise.  Fig. \ref{fig:ex} illustrates the
possibility of a ``full blind'' search in this situation. This means
that we perform the detection jointly with the estimation of the GW
chirp frequency and the source sky location.

\subsubsection{Description of the test signal}
Because of computational limitation, we restrict this study to rather
short chirps of $N=256$ samples, i.e., a chirp duration $T=250$ ms
assuming a sampling rate of $\nu_s=1024$ Hz. The chirp frequency follows
a random time evolution which however satisfies chirping rate
constraints. We make sure that the first and second derivatives of the
chirp frequency are not larger than $F'=9.2$ kHz/s and $F''= 1.57$
MHz/s$^2$ respectively. The chosen test signal has about $50$ cycles.
This is a larger number than
what is considered typically for burst GWs ( $\sim$ 10).

As a comparison with a well-known physical case, an inspiralling
(equal mass) binary with total mass $M=11 M_\odot$ reaches the same
maximum frequency variations at the last stable circular orbit.
(Binary chirps with larger total mass also satisfy these chirping rate
limits).

We set the SNR to $\rho=20$. The chirp is injected at the sky position
$\phi=2.8$ rad and $\theta=0.4$ rad where the contributions of
the individual detectors are comparable, namely the individual SNR are
$10.4$, $10.15$ and $13.77$ for Virgo, LIGO Hanford and LIGO
Livingstone respectively.

\subsubsection{Search parameters}
We search through the set of CCs defined over a TF grid with $N_t=128$
time intervals and $N=N_f=256$ frequency bins (using $f_s=2048$ Hz).
We set the regularity parameters to $N_r'=9$ and $N_r''=3$, consistent
to the above chirping rate limits.

We select an ad-hoc sky grid by dividing regularly the full range of
the source localization angles $\theta$ and $\phi$ into 128 bins. The
resulting grid has therefore a total of $N_\Omega = 16384$ bins. This
is probably much finer than is required to perform the detection
without missing candidate. However, this oversampling leads to precise
likelihood sky maps which helps to diagnose the method. With this
parameter choice, the estimated computational power required to
analyzed the data in real time is of $2.8$ TFlops, assuming an overlap
of $\mu=50 \%$ between successive data blocks. Because of the crude
choice for the sky grid, this requirement is probably over-estimated.

The result of the search is displayed in Fig. \ref{fig:ex} where we
see that the injection is recovered both in sky position and frequency
evolution. The source position is estimated at $\hat{\phi}=2.8$ rad and
$\hat{\theta}=0.39$ rad.


\begin{figure}
\includegraphics[width=\columnwidth]{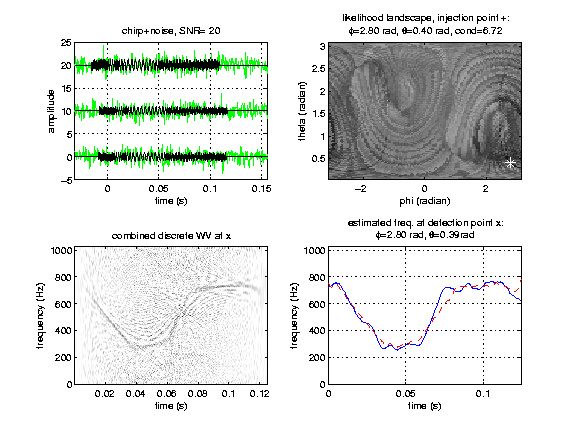}
\caption{\label{fig:ex} \textbf{Color online --- Coherent
    detection/estimation of a ``random'' chirp with network of three
    GW antennas}. In the data of three GW antennas (the two LIGO and
  Virgo), we inject (a) a ``random'' GW chirp emitted from a source at
  the position marked with a ``+'' at $\phi=2.8$ rad and $\theta=0.4$
  rad. We perform a full sky search using the best network chirplet
  chain algorithm. It produces a likelihood landscape (b) where we
  select the maximum.  This is the detection point and it is indicated
  with ``$\times$''. In (c) we show the combined WV distribution of
  the synthetic streams at the detection point. In (d), we compare the
  exact frequency of the chirp (solid/blue) with the estimation
  (dashed/red) obtained at the detection point.}
\end{figure}

\subsection{Regularized variants}
\label{sec:variants}
As shown in Sec. \ref{sec:snr_per_synth_streams}, the SNR carried by the synthetic
stream is proportional to the corresponding SV. When the GW polarization plane is
degenerate (i.e., when $\sigma_2$ is small), the second synthetic
stream contains almost only noise. We thus don't lose information if
we suppress its contribution from the statistic. This is the basic
idea of Klimenko \textit{et al.} in \cite{klimenko05:_const}.

We have seen that the estimation of the extrinsic parameters is an
ill-posed least-square problem in those cases.  Suppressing the
contribution of the second synthetic stream amounts to
\textit{regularizing} this problem \cite{klimenko05:_const}. In
practice, this regularization can be done in various ways,
corresponding to well-identified schemes.

A first possibility is to suppress the contribution of the second
synthetic stream when the conditioning number of $\vec{\Pi}$ is too
large (i.e., exceeds a given threshold). This scheme is referred to as
\textit{truncated SVD} \cite{neumaier98:_solvin_ill}. A second
possibility is to balance (divide) the contribution of the second
synthetic stream by the conditioning number. This is referred to as
the \textit{Tykhonov} approach \cite{neumaier98:_solvin_ill} and it
was proposed for regularizing burst searches in
\cite{rakhmanov06:_rank_tikhon}.

In Fig.~\ref{fig:reg}, we compare the likelihood landscape and
frequency estimate obtained with the standard statistic and its
regularized version using the Tykhonov approach. Visually, the
regularization improves the contrast and concentration of the
likelihood landscape around the injection point. This can be assessed
more quantitatively with the contrast defined as the ratio of the
likelihood landscape extremes. This contrast is improved by about $10
\%$ for the regularized statistic as compared to the standard version.
It is also interesting to compare the ``width'' of the detection peak
obtained with the two statistic. To do this, we measure the solid
angle of the the sky region where the statistic is larger than $90 \%$
of the maximum. This angle is reduced by a factor of $\sim$ 6 when
computed with the regularized statistic. There is however no major
improvement of the frequency estimate. More generally, it is unclear
whether the regularized statistic performs better than the standard
one.

\begin{figure}
\includegraphics[width=\columnwidth]{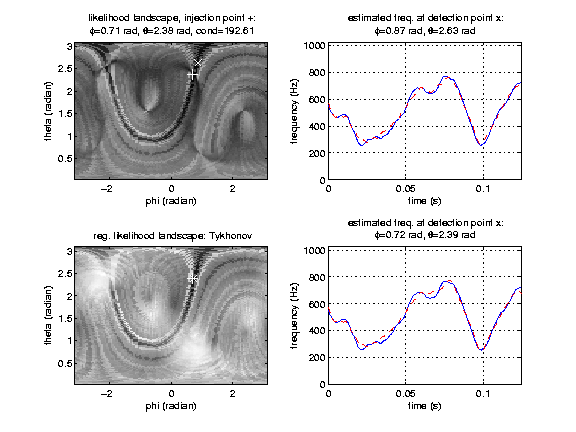}
\caption{\label{fig:reg} \textbf{Color online --- Coherent detection/estimation
    of a ``random'' chirp with network of three GW antennas. Standard
    and regularized statistic}. We compare the likelihood landscape
  (left) and frequency estimation (right) obtained using the standard
  (top) and Tykhonov-regularized (bottom) versions of the network
  statistic. The test signal is a ``random'' GW chirp injected at the
  sky location marked with a ``+''. This location has been chosen
  because of the associated large value of the conditioning number,
  namely $\cond(\vec{\Pi})\approx 192$.}
\end{figure}

\section{Concluding remarks}

The coherent detection of unmodeled chirps with a network of
GW have features and issues
in common with the burst one. In particular, the same geometrical
objects play a key role. While the noise spans the whole $d$
dimensional data space, GW signals (chirps or burst) only belong to a
two-dimensional (one dimension per GW polarizations) subspace, the
\textit{GW polarization plane}. Detecting GWs amounts to checking
whether the data has significant components in this plane or not.  To do so,
we compute the projections of the data onto a basis of the GW
polarization plane. In practice, this defines two instantaneous linear
mixtures of the individual detector data which we refer to as
\textit{synthetic streams}. Those may be considered as the output data
of ``virtual'' detectors. This combination is such that the GW
contributions from each real detector add constructively. The GW
signature thus has a larger amplitude in the synthetic streams while
the noise variance is kept at the same level.

The coherent detection amounts to looking for an excess in the signal
energy in one or both synthetic streams (depending on the GW
polarization model). This provides a generic and simple procedure to
produce a coherent detection pipeline from a one detector pipeline.
In the one detector case, the BCC search performs a path search in a
time-frequency distribution of the data. In the multiple detector
case, the BNCC search now use the joint time-frequency map obtained by
summing the time-frequency energy distributions of the two synthetic
streams. The approach does not restrict to chirp detection and it can
be applied to burst searches \cite{chatterji06:_coher}.

We demonstrated in a simplified situation that the full-sky blind
detection of an unmodeled chirp is feasible. This means that the
detection is performed jointly to estimate the source location and the
frequency evolution. The application of this method to the real data
however requires several improvements. First, the method has to be
adapted to the case where the detectors have different sensitivities.
In this respect, we already obtained first results \cite{rabaste07}.

We also have to refine the choice of the grid which samples the
celestial sphere. In the present work, we select an ad-hoc grid.
Clearly, the sky resolution and the bin shape depends on the geometry
of the sphere and the location and orientation of the detectors in the
considered network. A better grid choice (not too coarse to avoid SNR
loses, or and not too fine to avoid using useless computing resources)
should incorporate this information keeping the search performance
(detection probability and sky resolution) constant. In this respect,
we may consider to other parameterizations of the sky location which
makes the definition of the sky grid easier, for instance by choosing
the time-delays as investigated in \cite{Pai:2000zt}. We may also
explore hierarchical schemes for the reduction of the computational
cost.

The GW polarization plane depends on the detector antenna patterns
functions.  With the presently available networks, there are
significantly large sky regions where the antenna patterns are almost
aligned. In this case, the network observes essentially one
polarization and is almost insensitive to the other: the GW
polarization plane reduces to a one-dimensional space. The information
carried by the missing polarization lacks and this makes the
estimation of certain parameters ill-posed and hence very sensitive to
noise. We can evaluate that the variance of the estimate scales with
the condition number of the antenna pattern matrix.  When this number
(which quantifies in some sense the mutual alignment of the detectors)
is large, the estimation is ill-posed and we expect poor results.

This is an important issue for burst detection since it affects
significantly the shape of the estimated waveform (and, particularly
the regularity of its time evolution). This has motivated the
development of regularization schemes which penalize the estimation of
non-physical (i.e., irregular) waveforms. We have shown that this is
however less of a problem for chirps because of their more constrained
model.  Ill-conditioning only affects global scaling factors in the
chirp model.  Unlike bursts, no additional prior is available for
regularizing the estimation of these scaling factors.

The data space can be decomposed as the direct sum of the GW
polarization plane and its complementary. While GWs have zero
components in the latter null space, it is unlikely that instrumental
noise (including its non-Gaussian and non-stationary part) will. This
motivates the use of \textit{null streams} (i.e., the projection of
the data along a basis of the complementary space) to verify that a
trigger is indeed a GW candidate and not an instrumental artifact.
Since null streams are inexpensive to compute, we consider to use them
to make preemptive cuts in order to avoid the analysis of bad data.

\begin{acknowledgments}
  AP is supported by the Alexander von Humboldt Foundation's Sofja
  Kovalevskaja Programme (funded by the German Ministry of Education
  and Research). OR is supported by the Virgo-EGO Scientific Forum.
  The authors would like to thank J.-F. Cardoso for interesting
  exchange of ideas.
\end{acknowledgments}

\appendix
\section{Kronecker product: definition and properties}
\label{app:kron}
The Kronecker product $\otimes$ transforms two matrices $\vec{A} \in \C^{m \times
  n}$ and $\vec{B}\in \C^{p \times q}$ into the following matrix of $\C^{mp
\times nq}$ \cite{horn91:_topic_matrix_analy}
\be
\vec{A} \otimes \vec{B} \equiv  \left[ \ba{ccc}
a_{11} \vec{B} & \ldots & a_{1n} \vec{B}\\
\vdots & \ddots & \vdots \\
a_{m1} \vec{B} & \ldots & a_{mn} \vec{B}
\ea \right].
\ee

The Kronecker product is a linear transform and can be considered as a
special case of the tensor product. We define the operator $\stack{}$
to be the stack operator which transforms the matrix into a vector by
stacking its columns, {\it i.e.} $\vec{x}\equiv \stack{\vec{X}}$.  In
the text, we use the following property: 
\be
\label{eq:kron_stack}
(\vec{A} \otimes \vec{B}) \stack{\vec{X}} = \stack{\vec{B}\vec{X}\vec{A}^T}.
\ee

The proof of this property is straightforward.

\section{Interferometric detector response in terms of Gel'fand functions}
\label{app:detresp}
The GW response of a detector to an incoming GW can be obtained by computing the
interaction of the wave tensor $W$ with the detector tensor $D$ as follows
\footnote{Unless otherwise mentioned, the notations and symbols used in all appendices
are confined to those appendices only.}:
\be
s=\sum_{i,j=1}^3 W^{ij} D_{ij}.
\ee
The wave tensor is related to the incoming GW tensor in the TT 
gauge by $h_{ij} = 2 W_{ij}$. Both detector and wave tensors are rank 2
STF tensors. Any STF tensor can be expanded in the basis of
spin-weighted spherical harmonics of rank 2 and the rank-2 Gel'fand functions
provide the corresponding coefficients. Further, they are representation of
rotation group SO(3) and provide compact representation for the detector
response of any arbitrarily oriented and located detector on Earth which we
present in this appendix \cite{dhurandhar:1988}.

\paragraph{Wave tensor} --- The incoming GW tensor in TT gauge is given by
\be
h_{ij}^{TT} = (e_{xi} e_{xj} - e_{yi} e_{yj}) h_+ + 2 (e_{xi} e_{yj}) h_\times,
\ee
where ${\bf e_x}$ and ${\bf e_y}$ are unit vectors along the $x_w$ and
$y_w$ axes in the wave frame; $h_+$ and $h_\times$ are the
two GW polarizations. Let $\hat{m} = ({\bf e_x} + i {\bf e_y})/\sqrt{2}$
be a complex vector in the wave frame. Then, the wave tensor 
can be written down in terms of $m$ as
\be
\label{eq:Wij}
W^{ij} = \Re[(m^{i} m^{j})^* h] \,,
\ee
where we used the complex quantity $h=h_+ +i h_\times$ which combines
both GW polarizations.

The term $m_i m_j$ is a STF tensor of rank 2.
We choose to work in the detector frame for convenience.
Expanding $m_i m_j$  in terms of spin-weighted
spherical harmonics of rank 2, namely ${\cal Y}^{ij}_{2n}$ and the
rank-2 Gel'fand functions $T_{mn}(\phi'_e,\theta'_e,\psi'_e)$, we get
\be
\label{eq:mm}
m^i m^j = \sqrt{\frac{8 \pi}{15}} \sum_{n=-2}^2 {\cal Y}_{2n}^{ij} T_{2n} (\phi'_e,\theta'_e,\psi'_e) \,,
\ee
The angles $(\phi'_e,\theta'_e,\psi'_e)$ are the
Euler angles of the rotation operator which transforms the detector
coordinates into the wave coordinates.

Substituting Eq.~(\ref{eq:mm}) into Eq.~(\ref{eq:Wij}), we express
the wave tensor in terms of the Gel'fand functions as
\be
\label{eq:wij}
W^{ij} = \sqrt{\frac{8 \pi}{15}}~\sum_{n=-2}^2  \Re[h {\cal Y}_{2n}^{ij} T_{-2n}] \,.
\ee

\paragraph{Detector tensor} --- The detector tensor is
\be
\label{eq:dij}
D_{ij} = n_{1i} n_{1j} - n_{2i} n_{2j},
\ee
where $\vec{n}_1$ and $\vec{n}_2$ are the unit vectors along the first and second
arms of the interferometer. Recall that we choose the $x_d$-axis of the
detector frame along the bisector of the two arms. The $y_d$-axis is
chosen such that $(x_d,y_d,z_d)$ is a right handed coordinate system
with $z_d$ pointing towards the local zenith.  With this choice, we
have 
\be
\label{eq:det_ten}
D_{11}=D_{22}=0\,, \hspace{2cm} D_{12}=D_{21}= -1 \,.
\ee

From Eqs. (\ref{eq:dij}) and (\ref{eq:det_ten}), the detector response is
\be
\label{eq:detresp}
s=\Re[f^* h],
\ee
where the complex antenna pattern function is given by
\be
\label{eq:antpat2}
f=i [T_{2-2}(\phi'_e,\theta'_e,\psi'_e)-T_{22}(\phi'_e,\theta'_e,\psi'_e)] \,.
\ee

From the expansion of above Eq.~(\ref{eq:detresp}) in terms of GW
polarizations, it is consistent to define $f=f_+ + i f_\times$, which
yields Eq.~(\ref{eq:gw_resp}).

\paragraph{Extended complex antenna pattern for sources orbiting in a
  fixed plane } --- As discussed in Sec. \ref{sec:extendedAP}, the extended
antenna pattern functions incorporate the inclination angle $\e$ and
is given in Eq. (\ref{eq:ext_ant}) as
\be
\label{eq:antpat3}
{\tilde{f}} = \frac{1+\cos^2 \e }{2} f_{+} + i \cos \e f_{\times}\,.
\ee
where $f_{+,\times}$ depend on the relative orientation of wave frame
with respect to the detector frame. The detector to wave frame coordinate 
transformation can be split into two: detector to Earth's frame and Earth's
frame to wave frame by the following rotation transformations as given in Eqs.
(\ref{WtoE}) and (\ref{DtoE})
\be
\label{eq:rot_tras}
{\cal O}(\p'_e, \t'_e, \psi'_e) = {\cal O}(\p_e, \t_e, \psi_e) {\cal O}^{-1} (\a, \b, \g)\,.
\ee
The above successive rotation transformation can be translated into
the addition theorem of Gel'fand functions \cite{Gel:1963} as given below
\be
\label{eq:add}
T_{mn}(\phi'_e,\theta'_e,\psi'_e)=\sum_{l=-2}^{2} T_{ml}(\phi_e,\theta_e,\psi_e) T^*_{nl} (\a,\b,\g) \,.
\ee
We used the fact that the inverse rotation operator is associated to a complex conjugation.

Substituting in Eq.~(\ref{eq:antpat2}), we rewrite the antenna pattern functions in terms
of the Gel'fand functions as
\be
\label{eq:ant}
f = \sum_{s=-2}^2 i T_{2s} (\phi_e,\theta_e,\psi_e)[T_{2s}(\a,\b,\g)-T_{-2s}(\a,\b,\g)]^*.
\ee

Substituting in the extended beam pattern function given in
Eq.~(\ref{eq:antpat3}) and combining the dependencies upon $\psi=\psi_e$ and
$\e$, we get
\be
{\tilde{f}} = T_{22}(\psi,\e,0) d + T_{2-2} (\psi,\e,0) d^*,
\ee
where
\be
d= - \sum_{n=-2}^2 i T_{2 n}(\phi_e,\theta_e,0)[T_{2 n}(\a,\b,\g) - T_{-2 n} (\a,\b,\g)]^* \,.
\ee

There are various ways of expressing the antenna pattern functions. The main
advantages of this one is that it is particularly compact and that the
angles $\psi$ and $\epsilon$ get factorized from the rest of the
parameters. This helps in the maximization of the network LLR over the
extrinsic parameters.

\section{SVD of a two-column complex matrix}
\label{app:svd_rank2}
In this appendix, we obtain the SVD of a complex matrix of the type
$\vec{A} \equiv [\vec{a},\:\vec{a}^*]$ where
$\vec{a} \in \C^{N \times 1}$. The SVD decomposes $\vec{A}$ into the
product $\vec{A} = \vec{U}_A \vec{\Sigma}_A \vec{V}_A^H$ where
$\vec{U}_A$ and $\vec{V}_A$ are two orthogonal matrices and
$\vec{\Sigma}_A$ is a positive definite diagonal matrix. To obtain it
analytically, we first get the eigen-decomposition of 
\be
\vec{A}^H \vec{A} = \left[\ba{cc}\vec{a}^H \vec{a} & \vec{a}^H \vec{a}^*\\
  \vec{a}^T \vec{a}& \vec{a}^T \vec{a}^* \ea \right] \equiv \left[\ba{cc} a & b^* \\
  b & a \ea \right]. 
\label{eq:general_form}
\ee

We distinguish two cases depending on the value of $b$.

\subsection{For $|b|>0$}
The eigenvalues $\sigma_\pm$ and the eigenvectors $v_\pm$ of $\vec{A}^H \vec{A}$ 
are given below;
\bea
\sigma^2_\pm &=& a \pm |b|,\\
\vec{v}_\pm &=& \frac{1}{\sqrt{2}} \left[\ba{c} \exp (-i \varpi) \\ \pm 1 \ea \right] \,,
\eea
where $\varpi = \arg b =\arg[\vec{a}^T \vec{a}]$. The number of
non-zero eigenvalues of $\vec{A}^H \vec{A}$ gives the rank of
$\vec{A}$. The eigenvalues arranged in the descending order form the
diagonal values of $\vec{\Sigma}_A^2$ as shown below. The
eigenvectors $\vec{v}_\pm$ are the right-handed singular vectors of
$\vec{A}$ and form $\vec{V}_A$ as given below:
\bea
\vec{\Sigma}_A &=& \left[\ba{cc}{ \sigma_+} & 0\\
0& {\sigma_-}\ea \right]\\
\vec{V}_A &=& \left[\ba{cc} \vec{v}_+ & \vec{v}_- \ea 
\right]\,. 
\eea

We then form the matrix $\vec{U}_A = \vec{A} \vec{V}_A \vec{\Sigma}^{-1}_A \equiv
[\vec{u}_+,\: \vec{u}_-]$ containing the right-handed singular vectors, with 
\be
\label{eq:uA}
\vec{u}_{\pm} = \frac{\exp(-i \varpi) \vec{a} \pm \vec{a}^*} {\sqrt{2} \sigma_{\pm}}\,.   
\ee

The above expressions are valid only when the two SV are non zero. It
is possible that the smallest SV $\sigma_-$ vanishes. In this case,
the SVD collapses to $\vec{U}_A=\vec{u}_+$,
$\vec{\Sigma}_A=\sigma_+$, and $\vec{V}_A=\vec{v}_+$.

\subsection{For $|b| = 0$}
In this case, the matrix $\vec{A}^H \vec{A}=a \vec{I}_2$ is diagonal.
We thus have $\vec{\Sigma} = \sqrt{a} \vec{I}_2$, $\vec{V}_A =
\vec{I}_2$ and $\vec{U}_A = \vec{A}/\sqrt{a}$.

\bibliography{paper}

\end{document}